\documentclass[journal,onecolumn,12pt]{IEEEtran}
\usepackage[T1]{fontenc}
\usepackage{amsmath,amssymb,amsthm,mathtools}
\usepackage{xparse,scalerel}
\usepackage{graphicx,booktabs}
\usepackage{tikz}
\usetikzlibrary{decorations.markings,arrows.meta} 
\usepackage[caption=false,font=footnotesize,labelformat=simple]{subfig} 
\usepackage{xcolor}
\definecolor{darkblue}{rgb}{0,0,0.5}
\definecolor{darkgreen}{rgb}{0,0.5,0}
\usepackage{cite} 
\usepackage{array} 
\usepackage[colorlinks=true,linkcolor=darkblue,urlcolor=darkblue,citecolor=darkgreen]{hyperref} 
\hypersetup{pdftitle={Quantum Telatar formula: fading quantum capacity of Haar-random multimode attenuators},pdfauthor={Eric Culf and Victor V. Albert}}
\usepackage[nameinlink,capitalise]{cleveref}
\crefformat{equation}{(#2#1#3)}   
\Crefformat{equation}{Equation~(#2#1#3)}
\crefmultiformat{equation}{(#2#1#3)}{ and~(#2#1#3)}{, (#2#1#3)}{, and~(#2#1#3)}
\crefname{figure}{Fig.}{Figs.}
\Crefname{figure}{Fig.}{Figs.}
\crefname{subfigure}{Fig.}{Figs.}
\Crefname{subfigure}{Fig.}{Figs.}
\usepackage{etoolbox}

\theoremstyle{definition}
\newtheorem{definition}{Definition}
\theoremstyle{plain}
\newtheorem{theorem}[definition]{Theorem}
\newtheorem{lemma}[definition]{Lemma}
\newtheorem{corollary}[definition]{Corollary}
\theoremstyle{remark}
\newtheorem*{remark}{Remark}

\AtBeginEnvironment{theorem}{\crefalias{definition}{theorem}}
\AtBeginEnvironment{lemma}{\crefalias{definition}{lemma}}
\AtBeginEnvironment{corollary}{\crefalias{definition}{corollary}}

\newcommand{\mc}[1]{\mathcal{#1}}
\providecommand{\C}{}\renewcommand{\C}{\mathbb{C}}
\DeclareMathOperator{\Tr}{Tr}
\DeclarePairedDelimiter\parens{\lparen}{\rparen}
\DeclarePairedDelimiter\squ{[}{]}
\DeclarePairedDelimiter\ang{\langle}{\rangle}
\DeclarePairedDelimiter\abs{\lvert}{\rvert}
\DeclarePairedDelimiter\norm{\|}{\|}

\DeclareMathOperator*{\expec}{\scalerel*{\vphantom{\mathbb{E}}\text{\raisebox{0.3pt}{\scalebox{0.9}{$\mathbb{E}$}}}}{\sum}}
\DeclareDocumentCommand\ketbra{s m g}{
	\IfBooleanTF{#1}{
		\IfNoValueTF{#3}{
			\vphantom{#2}\left\lvert{#2}\middle\rangle\!\middle\langle{#2}\right\rvert
		}{
			\vphantom{#2#3}\left\lvert{#2}\middle\rangle\!\middle\langle{#3}\right\rvert
		}
	}{
		\IfNoValueTF{#3}{
			\vphantom{#2}\left\lvert\smash{#2}\middle\rangle\!\middle\langle\smash{#2}\right\rvert
		}{
			\vphantom{#2#3}\left\lvert\smash{#2}\middle\rangle\!\middle\langle\smash{#3}\right\rvert
		}
	}
}

\newcommand{\Ic}{I_{\mathrm c}}
\newcommand{\Fme}{\Phi^{\mathrm{me}}}
\newcommand{\dd}{\mathrm{d}}
\newcommand{\Ucal}{\mc{U}}
\newcommand{\id}{\mathrm{id}} 
\newcommand{\meanN}[1]{\langle\hat N\rangle_{#1}} 

\newcommand{\vva}[1]{{\ifmmode\else\leavevmode\fi\color{red}#1}} 
\newcommand{\nenv}{{n_{\mathrm{env}}}} 
\newcommand{\isoinput}[1]{\theta_{#1}}

\title{Quantum Telatar formula: fading quantum capacity of Haar-random multimode attenuators}
\author{Eric~Culf and Victor~V.~Albert%
\thanks{E.~Culf is with the Department of Applied Mathematics and the Institute for Quantum Computing, University of Waterloo, Waterloo, ON N2L 3G1, Canada.}%
\thanks{V.~V.~Albert is with the Joint Center for Quantum Information and Computer Science, NIST/University of Maryland, College Park, MD 20742 USA.}%
}
\begin{document}

\maketitle

\begin{abstract}

Cellular communication is done over an imperfectly predictable multimode environment, where the sender only knows about the distribution of possible intermode coupling and scattering a signal faces.
It may thus seem surprising that such communication is efficient.
Telatar proved that the sender's lack of exact knowledge about the channel, on average, does not significantly hinder multimode communication.
As the number $N$ of modes increases at fixed signal-to-noise ratio, the average capacity of a classical multimode fading channel grows linearly in $N$, like that of $N$ decoupled single-mode channels.
We study the quantum analogue in a Haar-random passive bosonic network and prove a similar scaling for unassisted quantum information transmission.
Suppose that the sender drives a fixed fraction of the input modes and the receiver collects more than half of the output modes.
Then the quantum capacity of such channels at zero temperature grows linearly with the number of modes at fixed photon number per mode, just like the classical capacity grows with the number of antennas.
At nonzero temperature, the Haar-averaged coherent information of the isotropic thermal input bounds the capacity from below, and its value per mode converges to an integral against Wachter's law.
We numerically identify positive achievable rates in models motivated by cryogenic microwave networks and room-temperature multimode fibers.

\end{abstract}

\begin{IEEEkeywords}
Bosonic channels, fading channels, Haar-random unitaries, multiple-input multiple-output (MIMO) systems, quantum capacity, Wachter's law.
\end{IEEEkeywords}

\section{Introduction}
\label{sec:intro}

\IEEEPARstart{C}{ellular} networks communicate through an environment that scatters each signal among many paths and antennas. The sender knows only the distribution of the resulting coupling between the modes, while the receiver learns each realization. In the classical literature, this communication scenario is known as a fading channel with channel-state information at the receiver~\cite{GoldsmithVaraiya1997,BiglieriProakisShamai1998}. 

It may seem surprising that such communication is efficient. The result that started multiple-input multiple-output (MIMO) communication says otherwise: at fixed signal-to-noise ratio, the capacity of a multiple-antenna link grows linearly with the number of antennas~\cite{Foschini1996,FoschiniGans1998}. Telatar's model makes this statement precise~\cite{Telatar1999}. A canonical MIMO channel is $y=Hx+w$ with an $R\times T$ Gaussian matrix $H$ that is redrawn on every use and known to the receiver, and the input $x$ obeys a power constraint. Its capacity is the average over $H$ of a log-determinant,
\begin{equation}
C=\expec_H\log\det\parens[\big]{I+\tfrac PTHH^\dagger}=\expec_H\sum_i\log\parens[\big]{1+\tfrac PT\lambda_i},
\label{eq:telatar}
\end{equation}
where the $\lambda_i$ are the eigenvalues of $HH^\dagger$. This capacity is attained by an isotropic Gaussian input. 
Telatar further showed that, when both numbers of antennas grow at a fixed ratio and fixed signal-to-noise ratio, the capacity per antenna converges to an integral against the Marchenko--Pastur law (\cref{tab:telatar}). The capacity thus grows linearly with the number of antennas, like that of decoupled single-antenna channels.
In other words, the sender's ignorance of $H$ does not, on average, hinder this growth.

We develop a quantum generalization of Telatar's MIMO channel and its corresponding capacity.
A passive optical network, including its loss modes, acts by a unitary matrix. We model its mixing by a Haar-random unitary and take the sender-to-receiver block as the transmission matrix.
\Cref{tab:telatar} summarizes the correspondence between Telatar's argument and ours.

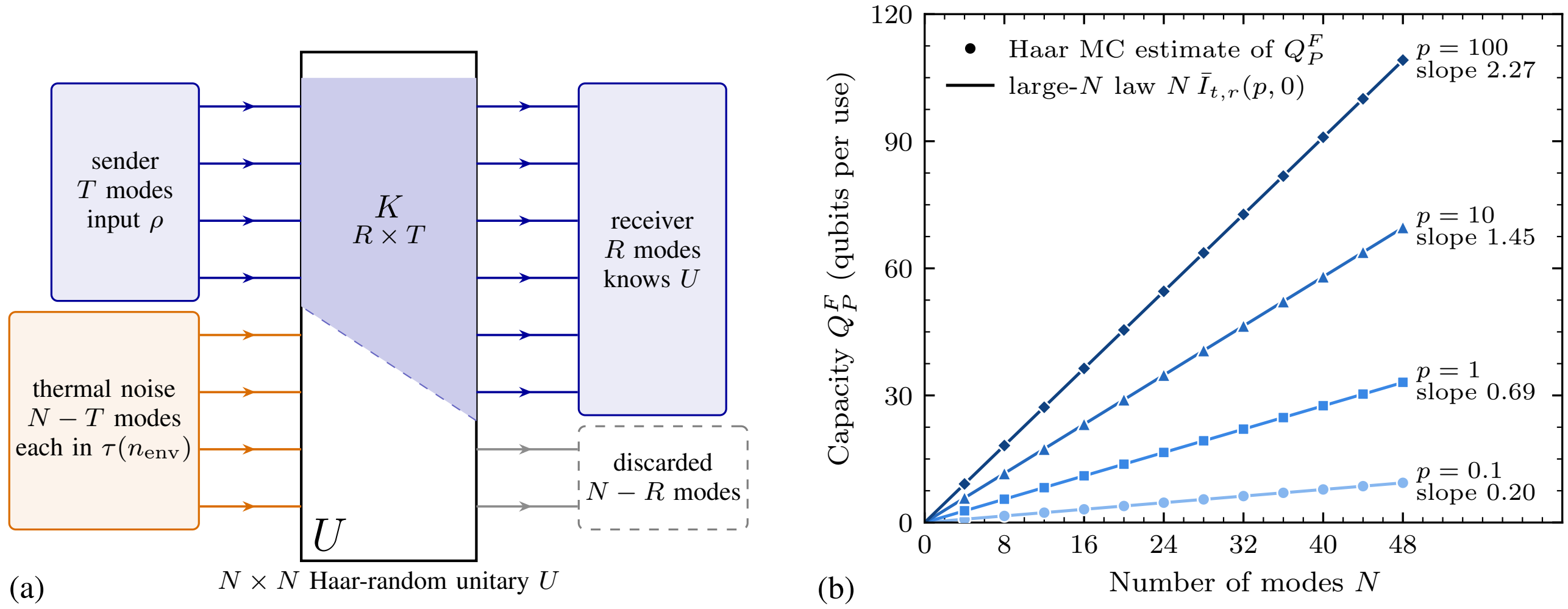
\begin{figure}[!t]
\centering
\subfloat{
\begin{tikzpicture}[
  font=\scriptsize,
  x=1cm, y=1cm,
  wire/.style={line width=0.7pt},
  flow/.style={postaction={decorate},
    decoration={markings, mark=at position 0.55 with {\arrow{Stealth[length=4pt,width=3.2pt]}}}},
  sig/.style={wire, flow, draw=sigblue},
  thermal/.style={wire, flow, draw=thermorange},
  discard/.style={wire, flow, draw=black!45},
  box/.style={rounded corners=2pt, line width=0.6pt, align=center, inner sep=2pt, minimum width=1.6cm},
  party/.style={box, draw=sigblue, fill=sigblue!8},
  noise/.style={box, draw=thermorange, fill=thermorange!8},
  sink/.style={box, draw=black!50, dashed},
]
\colorlet{sigblue}{blue!60!black}
\colorlet{thermorange}{orange!85!black}
\def\N{8}\def\T{4}\def\R{6}
\def\dy{0.62}           
\def\xs{1.0}            
\def\xb{2.1}            
\def\xc{4.0}            
\def\xr{5.1}            
\pgfmathsetmacro{\ytop}{\dy/2}
\pgfmathsetmacro{\ybot}{-(\N-1)*\dy-\dy/2}
\pgfmathsetmacro{\yT}{-(\T-1)*\dy-\dy/2}
\pgfmathsetmacro{\yR}{-(\R-1)*\dy-\dy/2}
\pgfmathsetmacro{\pad}{0.28}
\draw[line width=0.8pt, fill=white] (\xb,\ytop+\pad) rectangle (\xc,\ybot-\pad);
\fill[sigblue!20] (\xb,\ytop) -- (\xc,\ytop) -- (\xc,\yR) -- (\xb,\yT) -- cycle;
\draw[sigblue!60, line width=0.4pt, dashed] (\xc,\yR) -- (\xb,\yT);
\node[align=center] at ({(\xb+\xc)/2},{(\ytop+(\yT+\yR)/2)/2}) {{\small $K$}\\[-1pt]$R\times T$};
\node[anchor=south west, inner sep=3pt] at (\xb,\ybot-\pad) {\large $U$};
\node[anchor=north, align=center, inner sep=3pt] at ({(\xb+\xc)/2},\ybot-\pad)
  {$N\times N$ Haar-random unitary $U$};
\foreach \i in {1,...,\N} {
  \pgfmathsetmacro{\y}{-(\i-1)*\dy}
  \ifnum\i>\T
    \draw[thermal] (\xs,\y) -- (\xb,\y);
  \else
    \draw[sig] (\xs,\y) -- (\xb,\y);
  \fi
  \ifnum\i>\R
    \draw[discard] (\xc,\y) -- (\xr,\y);
  \else
    \draw[sig] (\xc,\y) -- (\xr,\y);
  \fi
}
\node[party, anchor=east, minimum height={(\T*\dy-0.12)*1cm}]
  at (\xs,{-(\T-1)*\dy/2}) {sender\\ $T$ modes\\ input $\rho$};
\node[noise, anchor=east, minimum height={((\N-\T)*\dy-0.12)*1cm}]
  at (\xs,{-(\N+\T-1)*\dy/2}) {thermal noise\\ $N-T$ modes\\ each in $\tau(\nenv)$};
\node[party, anchor=west, minimum height={(\R*\dy-0.12)*1cm}]
  at (\xr,{-(\R-1)*\dy/2}) {receiver\\ $R$ modes\\ knows $U$};
\node[sink, anchor=west, minimum height={((\N-\R)*\dy-0.12)*1cm}]
  at (\xr,{-(\N+\R-1)*\dy/2}) {discarded\\ $N-R$ modes};
\node[anchor=south west, inner sep=0, font=\small] at (current bounding box.south west) {(a)};
\end{tikzpicture}\label{fig:setting}}%
\hfil
\subfloat{%
\begin{tikzpicture}
\node[anchor=south west, inner sep=0] (plot) at (0,0) {\includegraphics[width=0.45\textwidth]{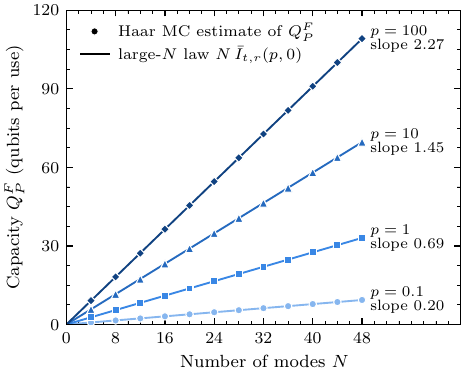}};
\node[anchor=south west, inner sep=0, font=\small] at (plot.south west) {(b)};
\end{tikzpicture}%
\label{fig:scalingplot}}
\caption{\label{fig:scaling}(a) \textsc{Quantum MIMO scenario.} $N$ bosonic modes pass through a passive linear-optical network, an $N\times N$ unitary $U$ drawn from the Haar measure on every use. A sender drives $T$ input modes with a state $\rho$, and the remaining $N-T$ inputs are each in the thermal state $\tau(\nenv)$. A receiver collects $R$ output modes and learns the realization $U$, and the remaining $N-R$ outputs are discarded. Blue wires mark the sender's inputs and the receiver's outputs, orange wires the thermal inputs, and gray wires the discarded outputs. The channel from sender to receiver, $\Phi_{K,\nenv}$ of \cref{def:channel}, depends on $U$ only through its top-left $R\times T$ block $K$ (shaded), whose squared singular values are the transmissivities $\eta_i$. The diagram is drawn for $N=8$ with $T=N/2$ and $R=3N/4$, the regime of the neighboring panel. (b) \textsc{Linear capacity scaling.} Fading quantum capacity of a Haar-random $N$-mode network at zero temperature versus the number of modes. The sender drives $N/2$ input modes with mean total photon number $P=pN$, that is, $2p$ photons per input mode, and the receiver collects $3N/4$ output modes and knows the realization. Markers show Monte Carlo estimates of the exact finite-$N$ capacity $Q^F_P=\bar I_N(P,0)$ of \cref{thm:capacity}, using $4\,000$ Haar unitaries (statistical error below the marker size). Lines show the large-$N$ law $N\,\bar I_{1/2,3/4}(p,0)$ of \cref{thm:limit}, whose slope $\bar I_{1/2,3/4}(p,0)$, an integral against Wachter's law, is given next to each line in qubits per mode. In this regime the large-$N$ law is exact: $N/4$ transmissivities equal $1$ and the joint distribution of the others is symmetric under $\eta\mapsto1-\eta$, so Haar averaging and the antisymmetry of $I_{2p,0}$ give $Q^F_P=N\,g(2p)/4$ at every $N$. At fixed photon number per mode the capacity grows linearly with the number of modes, the quantum counterpart of the linear growth with the number of antennas in Fig.~2 of Ref.~\cite{Telatar1999}.}
\end{figure}

Let $N$ modes pass through a passive linear network, an $N\times N$ unitary $U$ acting on the modes (\cref{fig:setting}). A sender drives $T$ of the input modes and a receiver collects $R$ of the output modes, so the channel between them is the top-left $R\times T$ block $K_U$ of $U$. 
With this Haar-distributed block in place of Telatar's Gaussian matrix, the classical model becomes the Jacobi MIMO channel of Dar, Feder, and Shtaif~\cite{DarFederShtaif2013}. Its transmissivities, the squared singular values of $K_U$, follow the Jacobi ensemble and, for many modes, Wachter's law~\cite{Wachter1980,Collins2005}. 
Ref.~\cite{KaradimitrakisMoustakasVivo2014} derived the large-$N$ ergodic mutual information of the Jacobi MIMO channel.


In the quantum setting, the sender's $T$ modes carry an arbitrary quantum state.
The receiver learns each realization of $U$, and we ask for the quantum capacity under a mean photon-number constraint. 
The motivation is the same as in the classical case. Optical fibers for space-division multiplexing carry independent classical or quantum signals in many spatial modes or cores~\cite{RichardsonFiniNelson2013,XavierLima2020}. Crosstalk between the modes or cores turns the fiber into a MIMO channel~\cite{WinzerFoschini2011}. Long multimode and coupled-core fibers can mix their modes strongly and at random~\cite{HoKahn2011,HayashiEtAl2022}. With negligible backscattering and intrinsic loss, the transmission matrix of such a fiber is modeled as a Haar-random unitary known only to the receiver~\cite{KaradimitrakisMoustakasVivo2014}. Modes that the sender and receiver leave unaddressed account for the loss: the channel between them is a corner of this unitary~\cite{DarFederShtaif2012,KaradimitrakisMoustakasVivo2014}. The transmission matrix decorrelates beyond a coherence bandwidth set by modal dispersion, so a much wider signal sees many nearly independent realizations and its outage capacity approaches the average capacity~\cite{DarFederShtaif2012,HoKahn2011Diversity}. Quantum light sees the same channel: a $55.3$~cm graded-index fiber supporting about $400$ modes at $810$~nm acts on it as a linear-optical network that mixes its modes nearly isotropically~\cite{LeedumrongwatthanakunEtAl2020}. Random unitary circuits in measurement-induced phase transitions provide a related theoretical setting (\cref{sec:related}). Cryogenic microwave networks offer another potential setting, as superconducting processors are now linked by cryogenic cables and waveguides over meters to tens of meters~\cite{MagnardEtAl2020,StorzEtAl2023,NiuEtAl2023}. Random-matrix models describe wave transport through chaotic microwave cavities~\cite{GradoniEtAl2014}, and residual thermal photons can affect superconducting circuits~\cite{KrinnerEtAl2019,WangEtAl2019}. These motivate our microwave example, which idealizes the transport as a Haar-random unitary.
Turbulent free-space links, the optical counterpart of the multiple-antenna wireless links in which MIMO originated~\cite{Foschini1996,FoschiniGans1998}, also exhibit mode mixing and fading~\cite{ChandrasekaranShapiroWang2014,Pirandola2021FreeSpace}. However, their high-loss regime makes them unsuitable settings for one-way quantum communication (\cref{sec:numerics}), with the two-way assisted capacity of our random multimode model left open (\cref{sec:twoway}).

Our key result is that the zero-temperature capacity grows linearly with the number of modes $N$ at fixed photon number per mode, as long as the receiver collects more than half of the output modes. At nonzero temperature our lower bound on the capacity grows in the same way wherever it is positive.
We derive this large-$N$ scaling analytically, and \cref{fig:scalingplot} compares Monte Carlo estimates of the exact finite-$N$ capacity with its large-$N$ law for $T=N/2$ and $R=3N/4$.

\Cref{sec:results} defines the fading capacity and states our results, with the zero-temperature case and the two-way assisted capacity in \cref{sec:zerotemp,sec:twoway}. \Cref{sec:numerics} numerically evaluates the rate for microwave and fiber models with representative intrinsic losses. \Cref{sec:related} relates the results to random circuits, to earlier work on fading bosonic channels, and to the finite-size Jacobi formula, and \cref{sec:conclusion} concludes. The appendices are self-contained and build up to the large-$N$ limit of the fading capacity. \Cref{app:onemode} treats the one-mode thermal attenuator and its coherent information $I_{n,\nenv}$. \Cref{app:channel} defines the multimode channel and the fading capacity, reduces the single-use coherent information of the isotropic input to a sum of one-mode terms, and states the capacity theorem (\cref{thm:capacity}). \Cref{app:limit} takes the large-$N$ limit by Wachter's law. \Cref{app:achievability} proves \cref{thm:capacity}(i), and \cref{app:exact} proves \cref{thm:capacity}(ii).

\begin{table}[!t]
\centering
\caption{\label{tab:telatar}Correspondence between Telatar's treatment of the Rayleigh MIMO channel~\cite{Telatar1999} and the present paper. Both columns use $T,R$ for the numbers of transmit and receive modes or antennas; $t,r$ denote the limiting mode fractions. Other symbols are defined in \cref{tab:symbols}.}
\renewcommand{\arraystretch}{1.3}
\begin{tabular}{>{\raggedright\arraybackslash}p{0.42\textwidth}|>{\raggedright\arraybackslash}p{0.5\textwidth}}
\hline
Telatar~\cite{Telatar1999} & This paper \\
\hline\hline
$y=Hx+w$ with $H$ an $R\times T$ Gaussian matrix & $\Phi_U=\Phi_{K_U,\nenv}$ with $K_U$ the $R\times T$ corner of a Haar $U\in U(N)$ \\
\hline
receiver knows $H$, sender knows its distribution & receiver knows $U$, sender knows the Haar measure \\
\hline
power $\expec\norm{x}^2\le P$ & mean photon number $\le P$ per use \\
\hline
$\log\det(I+\tfrac{P}{T}HH^\dagger)=\sum_i\log(1+\tfrac{P}{T}\lambda_i)$ & $\sum_i I_{P/T,\nenv}(\eta_i)$ with $\eta_i\in\operatorname{spec}(K_U^\dagger K_U)$ \\
\hline
isotropic Gaussian input optimal, by concavity and $HV\sim H$ & isotropic thermal input optimal at $\nenv=0$ and $R\ge N/2$, by \cref{lem:averaged,lem:tangent} \\
\hline
$C=\expec\log\det(\cdot)$ exactly & $Q^F_P=\bar I_N(P,0)$ for $R\ge N/2$, $Q^F_P=0$ for $R\le N/2$, and $Q^F_P\ge \bar I_N(P,\nenv)$ for $\nenv>0$ (\cref{thm:capacity}) \\
\hline
eigenvalues $\lambda_i$ of $HH^\dagger$: Wishart ensemble & transmissivities $\eta_i$: Jacobi ensemble~\cite{Collins2005,SimonMoustakas2006} \\
\hline
$C=\int_0^\infty\log(1+\tfrac{P}{T}\lambda)\,\mathcal L(\lambda,\lambda)\,\dd\lambda$, with $\mathcal L$ the weighted Christoffel--Darboux kernel of the first $\min\{T,R\}$ orthonormal Laguerre polynomials for the weight $\lambda^{|T-R|}e^{-\lambda}$ & For $T\le R$ and $T+R\le N$, $\bar I_N(P,\nenv)=\int_0^1I_{P/T,\nenv}(x)\,\mathcal K_N(x,x)\,\dd x$, with $\mathcal K_N$ the weighted Christoffel--Darboux kernel of the first $T$ orthonormal Jacobi polynomials for the weight $x^{R-T}(1-x)^{N-R-T}$, see \cref{eq:jacobi} \\
\hline
Marchenko--Pastur limit & Wachter limit $\nu_{t,r}$ \\
\hline
\end{tabular}
\end{table}

\section{Results}
\label{sec:results}

Consider $N$ bosonic modes passing through the network of \cref{fig:setting}, an $N\times N$ unitary $U$ acting on the mode operators. A sender controls $T$ of the input modes, a receiver collects $R$ of the output modes, and the other $N-T$ inputs carry thermal noise with mean photon number $\nenv$ per mode. For a fixed and known $U$ this is a multimode thermal attenuator, a Gaussian channel whose capacities are treated in Refs.~\cite{HolevoWerner2001,Holevo2019}. It depends on $U$ only through the block $K_U$ (\cref{lem:charfun}), and its transmissivities $\eta_1(U),\dots,\eta_T(U)$ are the eigenvalues of $K_U^\dagger K_U$. For $\nenv>0$ its quantum capacity is not known in general, even for one mode~\cite{RosatiMariGiovannetti2018,NohAlbertJiang2019}. We instead draw $U$ from the Haar measure independently on every use and let the receiver, but not the sender, learn each realization. 

A fading code of block length $m$ consists of an encoder and a family of decoders indexed by the $m$ realizations (\cref{def:code}). Its mean photon number is at most $mP$, and its error is the entanglement-transmission error averaged over the realizations. The fading capacity $Q^F_P$ is the supremum of the achievable rates. No classical communication between sender and receiver is allowed, and we therefore call $Q^F_P$ the one-way capacity, in contrast with the capacity assisted by two-way classical communication (\cref{sec:twoway}).

Write $\tau(n)$ for the one-mode thermal state with mean photon number $n$ and $\isoinput P=\tau(P/T)^{\otimes T}$ for the isotropic thermal input, which puts $P/T$ photons in each sender mode. Write also $I_{n,\nenv}(\eta)$ for the coherent information of the one-mode thermal attenuator of transmissivity $\eta$ with input $\tau(n)$, a closed-form function~\cite{HolevoWerner2001,NohPirandolaJiang2020} recalled in \cref{lem:onemode}. For every realization, the coherent information of $\isoinput P$ through the channel is a sum of one-mode terms (\cref{thm:singleuse}),
\begin{equation}
\Ic\parens[\big]{\isoinput P,\Phi_U}=\sum_{i=1}^{T}I_{P/T,\nenv}\parens[\big]{\eta_i(U)}.
\label{eq:sum}
\end{equation}
Its Haar average is an achievable rate at every temperature (\cref{thm:capacity}(i)),
\begin{equation}
Q^F_P\ \ge\ \bar I_N(P,\nenv)=\int\dd U\sum_{i=1}^{T}I_{P/T,\nenv}\parens[\big]{\eta_i(U)},
\label{eq:rate}
\end{equation}
and at $\nenv=0$ the bound is the capacity for $R\ge N/2$, while the capacity vanishes for $R\le N/2$ (\cref{thm:capacity}(ii)):
\begin{equation}
Q^F_P=\begin{cases}\bar I_N(P,0),&R\ge N/2,\\0,&R\le N/2\end{cases}\qquad(\nenv=0).
\label{eq:capacity}
\end{equation}

\Cref{eq:rate} is the quantum counterpart of Telatar's capacity formula \cref{eq:telatar}: the one-mode coherent information replaces $\log(1+\tfrac PT\lambda)$, the transmissivities replace the eigenvalues of $HH^\dagger$, and the Haar average replaces the Gaussian average.

By \cref{lem:charfun}(b), the thermal channel is the vacuum channel followed by the classical Gaussian noise $\mc N_{K_U,\nenv}$ on the output. The receiver knows $U$ and can add this noise to the output of the vacuum channel. Thus every fading code for $\nenv>0$ is a fading code for $\nenv=0$ with the same error, and, writing $Q^F_P(\nenv)$ for the capacity at environment occupation $\nenv$, $Q^F_P(\nenv)\le Q^F_P(0)$. This has two consequences. First, the capacity is zero whenever the receiver collects at most half of the output modes: $Q^F_P=0$ for $R\le N/2$ at every $\nenv$. Second, $I_{n,\nenv}\le I_{n,0}$ pointwise and $\bar I_N(P,\nenv)\le\bar I_N(P,0)$, by data processing of coherent information under the same output channel. Hence, for $R\ge N/2$, the capacity at every temperature is sandwiched between the achievable rate and the zero-temperature capacity,
\begin{equation}
\bar I_N(P,\nenv)\ \le\ Q^F_P(\nenv)\ \le\ Q^F_P(0)=\bar I_N(P,0)\qquad(R\ge N/2).
\label{eq:sandwich}
\end{equation}

Mimicking Telatar, we develop a large-$N$ limit of the above capacity formula. Let $N\to\infty$ with $T/N\to t$, $R/N\to r$, and $P/N\to p$, at fixed photon number $p$ per mode. Then
\begin{equation}
\frac{\bar I_N(P,\nenv)}{N}\to\bar I_{t,r}(p,\nenv)=\int I_{p/t,\nenv}\,\dd\nu_{t,r}
\end{equation}
(\cref{thm:limit}), where $\nu_{t,r}$ is Wachter's law, the limiting law of the transmissivities of a Haar-random corner~\cite{Wachter1980,Collins2005}. It consists of a density on the interval $[\alpha_-,\alpha_+]$ plus atoms at $0$ and $1$,
\begin{align}
\nu_{t,r}&=w_{t,r}(x)\,\dd x+(t-r)_+\,\delta_0+(t+r-1)_+\,\delta_1,\label{eq:wachterlaw}\\
w_{t,r}(x)&=\frac{\sqrt{(\alpha_+-x)(x-\alpha_-)}}{2\pi x(1-x)}\,\mathbf 1_{[\alpha_-,\alpha_+]}(x),\notag
\end{align}
with $\alpha_\pm=(\sqrt{t(1-r)}\pm\sqrt{r(1-t)})^2$. The atoms are exact already at finite $N$: for almost every $U$, exactly $(T-R)_+$ transmissivities vanish and $(T+R-N)_+$ equal $1$ (\cref{app:limit}). Here $I_{n,\nenv}(0)=-g(n)$ and $I_{n,\nenv}(1)=g(n)$, where $g(x)=(x+1)\log_2(x+1)-x\log_2x$ is the entropy of a thermal mode with mean photon number $x$. Splitting the integral into bulk and atoms therefore gives
\begin{equation}
\bar I_{t,r}(p,\nenv)=\int_{\alpha_-}^{\alpha_+}I_{p/t,\nenv}(x)\,w_{t,r}(x)\,\dd x+\squ[\big]{(t+r-1)_+-(t-r)_+}\,g(p/t).
\label{eq:limit}
\end{equation} Consequently $\liminf_NQ^F_{P_N}/N\ge\bar I_{t,r}(p,\nenv)$ at every temperature, and at $\nenv=0$ and $r\ge\frac12$ the capacity per mode converges to $\bar I_{t,r}(p,0)$ (\cref{cor:permode}). Where the limiting achievable rate is positive, the capacity grows linearly with the number of modes at fixed photon number per mode, the quantum counterpart of the linear growth with the number of antennas in Telatar's theorem.

In the language of information theory, our results are three statements. First, \cref{thm:capacity}(i) is a coding theorem: the Haar-averaged coherent information of the isotropic thermal input is achievable under a mean photon-number constraint, at every temperature and for all $T$ and $R$. Second, \cref{thm:capacity}(ii) is a single-letter converse at zero temperature. For $R\ge N/2$, no input, entangled across uses or not, has larger averaged coherent information than the isotropic thermal input of the same photon number, so the fading capacity is $\bar I_N(P,0)$ exactly. For $R\le N/2$, the averaged coherent information of every input is nonpositive, so the capacity is zero. Third, \cref{thm:limit} and \cref{cor:permode} are the asymptotics: the rate per mode converges to an integral against Wachter's law, which is the capacity per mode at zero temperature and $r\ge\frac12$. 


\subsection{Zero-temperature case}
\label{sec:zerotemp}

At $\nenv=0$ the one-mode function reduces to
\begin{equation}
I_{n,0}(\eta)=g(\eta n)-g\parens[\big]{(1-\eta)n},
\label{eq:zerotemp}
\end{equation}
which is antisymmetric about $\eta=\frac12$. The fading capacity is then known exactly: $\bar I_N(P,0)$ for $R\ge N/2$ and zero for $R\le N/2$. This zero-temperature capacity bounds the capacity at every temperature from above, while the rate $\bar I_N(P,\nenv)$ bounds it from below. \Cref{sec:numerics} evaluates the extension of this lower bound to two models with representative intrinsic losses. 

Two limiting cases are instructive. First, when the sender drives half of the inputs, $t=\frac12$, the bulk of Wachter's law is symmetric about $\eta=\frac12$, and the antisymmetry of $I_{n,0}$ makes this bulk contribute nothing. For $r\ge\frac12$ the zero-temperature rate per mode is therefore
\begin{equation}
\bar I_{1/2,r}(p,0)=(r-\tfrac12)\,g(2p),
\label{eq:halfinput}
\end{equation}
carried entirely by the $(r-\frac12)N$ perfectly transmitted modes. In the regime of \cref{fig:scalingplot}, $T=N/2$ and $R=3N/4$, the joint distribution of the nontrivial transmissivities has the same symmetry at every $N$, so $Q^F_P=N\,g(2p)/4$ exactly. 

Second, the model so far has no loss other than the uncollected modes. Real links also have propagation and component losses. We add a uniform intrinsic transmissivity $\eta_f\in[0,1]$ on every mode, with the same environmental occupation $\nenv$, and ask how far the results carry over. Every transmission eigenvalue is then below one when $\eta_f<1$. The answer is a hard limit on one-way communication through such a network. At least $10\log_{10}2\approx3$~dB of intrinsic loss, about $15$~km if the propagation attenuation is $0.2$~dB/km and other losses are neglected, makes the one-way capacity vanish for every $N$, as it does for one mode.

Three facts give this limit. Thermal attenuators at the same temperature compose, $\Phi_{K_2,\nenv}\circ\Phi_{K_1,\nenv}=\Phi_{K_2K_1,\nenv}$ by substitution into \cref{eq:charfun}, so the loss turns $\Phi_U$ into $\Phi_{\sqrt{\eta_f}K_U,\nenv}$ whether it acts before or after the network, and the transmissivities become $\eta_f\eta_i(U)$. The proofs of \cref{thm:singleuse} and \cref{thm:capacity}(i) use only that $K_U$ is a contraction depending continuously on $U$, so the rate formulas hold with $\eta_i$ replaced by $\eta_f\eta_i$. \Cref{lem:degrade} does not transfer, and \cref{thm:capacity}(ii) is not claimed for $\eta_f<1$.

To distinguish this extension from the unattenuated model, write $\bar I_N(P,\nenv;\eta_f)$ for \cref{eq:rate} with $I_{P/T,\nenv}(\eta_i)$ replaced by $I_{P/T,\nenv}(\eta_f\eta_i)$. Its large-$N$ limit is
\begin{align}
\bar I_{t,r}(p,\nenv;\eta_f)
&=\int I_{p/t,\nenv}(\eta_f x)\,\dd\nu_{t,r}(x)\notag\\
&=\int_{\alpha_-}^{\alpha_+}I_{p/t,\nenv}(\eta_f x)w_{t,r}(x)\,\dd x
 +(t+r-1)_+ I_{p/t,\nenv}(\eta_f)-(t-r)_+g(p/t).
\label{eq:lossrate}
\end{align}
The proof of \cref{thm:limit} applies to this continuous integrand. The atom at one moves to $\eta_f$, while the atom at zero is unchanged. Omitting the final argument still means $\eta_f=1$.

 For $\eta_f\le\frac12$ the capacity nevertheless vanishes at every temperature. A one-mode attenuator of transmissivity $\eta\le\frac12$ is antidegradable~\cite{CarusoGiovannetti2006,WolfPerezGarciaGiedke2007} at every temperature: the reflected mode of its beam splitter carries $\Phi_{\sqrt{1-\eta},\nenv}(\rho)$, and the composition rule gives $\Phi_{\sqrt{\eta/(1-\eta)},\nenv}\circ\Phi_{\sqrt{1-\eta},\nenv}=\Phi_{\sqrt\eta,\nenv}$. Following the loss by $\Phi_U$ preserves antidegradability for each fixed $U$. Thus the coherent information of every input is nonpositive for every realization, and the first four steps of the proof of \cref{thm:capacity}(ii) give $Q^F_P=0$, as in the case $R\le N/2$. The threshold $\frac12$ of the one-mode attenuator~\cite{CarusoGiovannetti2006,WolfPerezGarciaGiedke2007} therefore governs the whole network. The lossy modes still carry something, however: \cref{sec:twoway} records what two-way assistance recovers from them.

\subsection{Towards two-way capacities}
\label{sec:twoway}

The two-way assisted capacity is the quantum capacity of our fading channel when sender and receiver may also exchange classical messages freely in both directions. With this assistance, quantum communication and entanglement distribution have the same capacity, so we quote it in ebits per use~\cite{PirandolaEtAl2017}. It is never smaller than the one-way capacity $Q^F_P$. We use our one-way results to bound the contribution of the lossy modes from above at zero temperature, and propose a candidate lower bound under the energy constraint.

The lossy modes contribute nothing to the zero-temperature one-way rate at $t=\frac12$, but with two-way classical assistance a pure-loss mode of transmissivity $\eta$ carries up to $-\log_2(1-\eta)$ ebits per use~\cite{PirandolaEtAl2017}, averaged over the fading~\cite{Pirandola2021FreeSpace}. The derivation parallels the one-way one, and we only outline it. Summing this bound over the transmissivities of a realization and averaging over the Haar measure gives an expression of the form of \cref{eq:rate}, with $-\log_2(1-x)$ in place of $I_{P/T,\nenv}(x)$. It bounds the two-way rate from above, since a sender who also learned $U$ could do no worse. The large-$N$ limit of \cref{thm:limit} then turns it into an integral against Wachter's law. The atom of Wachter's law at $\eta=1$ is excluded, since a perfectly transmitted mode has unbounded two-way capacity without an energy constraint, and the atom at $\eta=0$ contributes nothing. The bulk integral
\begin{equation}
\int_{\alpha_-}^{\alpha_+}\squ[\big]{-\log_2(1-x)}\,w_{1/2,r}(x)\,\dd x,
\label{eq:twowaybulk}
\end{equation}
evaluated numerically, is $0.311$ ebits per mode at $r=\frac34$ and $0.610$ at $r=0.4$. At $r=0.4$ the one-way capacity $Q^F_P$ is zero at every temperature (\cref{thm:capacity}), so classical assistance is the only route to a positive quantum rate there.

Under a mean photon-number constraint, a candidate lower bound on the two-way assisted capacity comes from reverse hashing of the isotropic two-mode-squeezed input. Reverse hashing distills entanglement at the rate of the reverse coherent information when classical communication from the receiver is free.
Reverse hashing then gives, at every $R$, the candidate rate
\begin{equation}
\int\dd U\sum_{i=1}^{T}\squ[\Big]{g\parens[\big]{\tfrac PT}-g\parens[\big]{(1-\eta_i(U))\tfrac PT}}.
\label{eq:twowayrate}
\end{equation}
However, the achievability argument of \cref{app:achievability} does not transfer as it stands to this quantity. 
That argument rests on two ingredients that fail for the reverse coherent information. The unassisted coding theorem of Ref.~\cite{WildeQi2018} does not cover reverse hashing, and the flag entropy cancels in the coherent information of the flagged channel but not in its reverse coherent information. A proof would need two further ingredients that we do not provide: an energy-constrained reverse hashing theorem, and a receiver who announces the partition cell of each realization. The energy-constrained converses of Refs.~\cite{FalcoMeleLamiGiovannetti2026,Pirandola2026TwoWay,PirandolaSpedalieri2026Multimode,MishraWilde2026Gaussian} are the natural tool for a matching upper bound, and we leave the two-way assisted capacities of this channel to future work.

\section{Numerical examples}
\label{sec:numerics}

\begin{figure}[!t]
\centering
\input{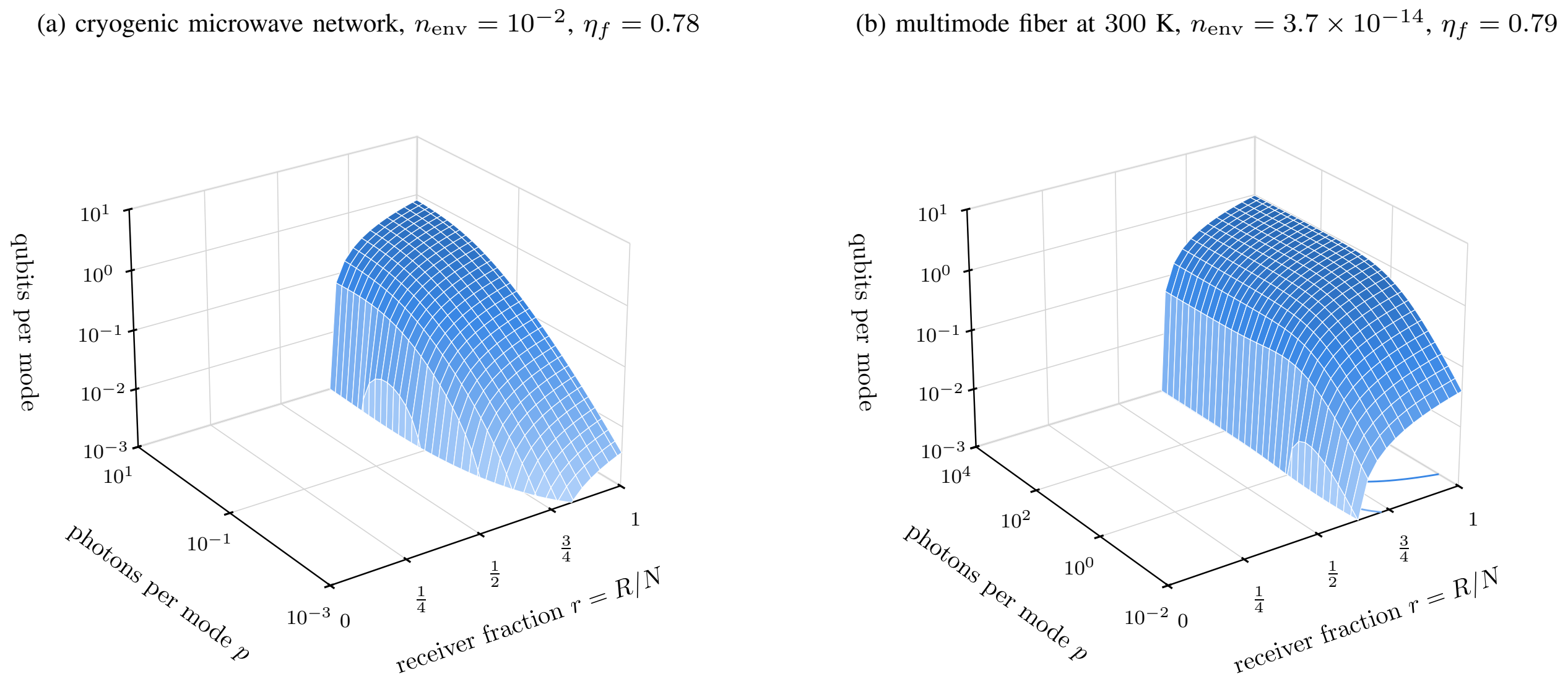}
\caption{\label{fig:thermal}Lower bound on the fading quantum capacity per mode at nonzero temperature, for two settings with a uniform intrinsic transmissivity $\eta_f$ on every mode (\cref{sec:zerotemp}). Each panel shows $\bar I_{t,r}(p,\nenv;\eta_f)$ of \cref{eq:lossrate}, the large-$N$ limit of $\bar I_N(P,\nenv;\eta_f)/N$. The axes are the receiver fraction $r=R/N$ and the photon number per mode $p=P/N$, the sender drives half the inputs ($t=\tfrac12$), and $\nenv$ is the mean background photon number per mode. The extension of \cref{thm:capacity}(i) described in \cref{sec:zerotemp} makes the surface a lower bound on the capacity per mode. With $\eta_f<1$ it is a lower bound even at $\nenv=0$ (\cref{sec:zerotemp}). Only rates between $10^{-3}$ and $10$ qubits per mode are shown; the visible edge at the floor is not a zero-rate contour. For $r\le\tfrac12$ the capacity vanishes at every temperature, so the bound is nonpositive (\cref{sec:results}). Lines on the floor are contours of the surface at one decade spacing, and shading encodes the height. The $p$ axis and the vertical axis are logarithmic. (a) \textsc{Cryogenic microwave network.} The environmental modes have residual occupation $\nenv=10^{-2}$, that of a $5$~GHz mode at $52$~mK, and $\eta_f=0.777$ uses the photon transmission of a cryogenic link as a uniform-loss benchmark (\cref{sec:numerics}). At $r=\tfrac34$ the bound is $49\%$ of its zero-temperature value at $p=10^{-2}$ and $85\%$ at $p=1$. It is positive only above $r\approx0.756$ at $p=10^{-3}$, $r\approx0.699$ at $p=10^{-2}$, and $r\approx0.669$ at $p=1$, against $r\approx0.645$, $0.646$, and $0.654$, respectively, at zero temperature with the same intrinsic loss. At $r=1$ it reaches $0.81$ qubits per mode at $p=10$. A nonpositive bound for $r>\tfrac12$ does not establish zero capacity. (b) \textsc{Multimode fiber at $300$~K.} Bright signals, with $\nenv$ the thermal occupation of a mode at $1550$~nm and $\eta_f=10^{-0.1}\approx0.794$ a representative $1$~dB propagation loss, corresponding to $5$~km at $0.2$~dB/km (\cref{sec:numerics}). The computed thermal correction is below $10^{-12}$ qubits per mode throughout the plotted grid. The bound is positive above $r\approx0.63$ at $p=10^{-2}$ and $r\approx0.66$ at $p=10^{4}$, and it saturates at $0.97$ qubits per mode at $r=1$.}
\end{figure}

\Cref{fig:thermal} evaluates $\bar I_{t,r}(p,\nenv;\eta_f)$ of \cref{eq:lossrate} for microwave and fiber models with representative intrinsic losses. For the microwave example we assign $\nenv=10^{-2}$ to every environmental input mode, the occupation of a $5$~GHz mode at $52$~mK. This representative choice lies within the range $6\times10^{-4}$ to $0.15$ surveyed for microwave readout cavities in Ref.~\cite{WangEtAl2019}. Single-photon microwave links~\cite{MagnardEtAl2020} motivate exploring small input occupations: the panel covers $10^{-3}\le p=P/N\le10$, or $2\times10^{-3}$ to $20$ photons per sender mode because $t=\tfrac12$. The measured noise levels motivate our parameters; a common thermal occupation for the unaddressed inputs and the intrinsic-loss environment, uniform attenuation across modes, and independent Haar mixing on each use remain modeling assumptions. Perfect knowledge of each realization at the receiver is also assumed; channel-estimation overhead is not included. For the fiber, the thermal occupation of a mode at $1550$~nm and $300$~K is $3.65\times10^{-14}$.

Cryogenic microwave links can operate below the $3$~dB loss threshold of \cref{sec:zerotemp}. Ref.~\cite{MagnardEtAl2020} reports $22.3\%$ photon loss (about $1.10$~dB), dominated by a circulator, even though the waveguide attenuation is below $1$~dB/km. Panel (a) uses $\eta_f=1-0.223=0.777$ as a benchmark for propagation and insertion loss. The experiment transfers photons at $8.406$~GHz and separately reports absorption inefficiency and qubit decay. Combining its photon-loss value with our representative $5$~GHz noise scale and applying it uniformly to a Haar network is an illustrative model.

For panel (b), we choose $5$~km and $0.2$~dB/km, giving $\eta_f=10^{-0.2\times5/10}\approx0.794$. This attenuation scale is consistent with low-loss coupled-core fiber: Ref.~\cite{HayashiEtAl2017LowLoss} reports $0.158$~dB/km at $1550$~nm in a coupled four-core fiber. The chosen length and uniform attenuation are assumptions; extra coupling, connector, and mode-dependent losses are omitted. The $400$-mode experiment of Ref.~\cite{LeedumrongwatthanakunEtAl2020} instead used $55.3$~cm at $810$~nm and does not establish these telecom-wavelength parameters for $400$ modes. With intrinsic loss the surfaces are lower bounds even at zero temperature, since \cref{thm:capacity}(ii) is not claimed for $\eta_f<1$ (\cref{sec:zerotemp}).

For the microwave example at $r=\tfrac34$, the lower bound of \cref{thm:capacity}(i) is $49\%$ of its zero-temperature value at $p=10^{-2}$ and $85\%$ at $p=1$. The intrinsic loss moves the positivity threshold of the bound well above $r=\tfrac12$. At $p=10^{-2}$ its zero-temperature threshold is $r\approx0.646$, and thermal noise raises this to $r\approx0.699$. At $p=10^{-3}$ the corresponding thresholds are $0.645$ and $0.756$ (\cref{fig:thermal}). These are thresholds of this achievable rate. With intrinsic loss, a nonpositive lower bound does not determine the capacity threshold even at zero temperature.

For the room-temperature fiber the correction is negligible throughout the bright range of the figure. The scale of the rate is set by the number of modes. Without intrinsic loss, at zero temperature, with $t=\tfrac12$, $r=\tfrac34$, and one photon per mode, the capacity is $N\,g(2)/4\approx0.689N$ qubits per use, or $275.5$ qubits per use for the ideal model with $N=400$. With the loss of panel (b), the large-$N$ achievable-rate estimate is $0.20025N$, approximately $80.1$ qubits per use at $N=400$. A single known mode of transmissivity $\tfrac34$ carrying the same $400$ photons achieves $1.58$ qubits per use~\cite{WildeHaydenGuha2012,WildeQi2018}, below its unconstrained limit $\log_23$~\cite{WolfPerezGarciaGiedke2007}. At zero temperature, a single mode with the same intrinsic loss, of transmissivity $(3/4)\eta_f\approx0.596$, achieves $0.558$ qubits per use. Without intrinsic loss, at zero temperature in the photon-starved regime, the rate per photon is $(2r-1)\,g(2p)/(2p)$ for $t=\tfrac12$ and $r\ge\tfrac12$, which gives $6.9$ qubits per photon at $p=10^{-4}$ and $10.2$ at $p=10^{-6}$ for $r=\tfrac34$. A single known mode of transmissivity $\tfrac34$ gives $6.7$ and $10.0$ at the same photon number per sender mode. With the loss of panel (b), the multimode rate per photon is $2.6$ and $3.8$ at these two points, and a single mode of transmissivity $0.596$ gives the same two values. Thus the multimode channel retains high photon efficiency under fading, while the gain over this single-mode benchmark is in the rate per use. For fixed intrinsic loss, a positive limiting achievable rate still implies linear growth with $N$. Its growth with $p$ is bounded: for $t=\tfrac12$, $r\ge\tfrac12$, and $\tfrac12<\eta_f<1$, monotonicity of $I_{n,0}(\eta)$ in $\eta$ makes the bulk contribution nonpositive, and
\begin{equation}
\bar I_{1/2,r}(p,0;\eta_f)\le(r-\tfrac12)\log_2\frac{\eta_f}{1-\eta_f}.
\label{eq:ratecap}
\end{equation}
This bounds the plotted achievable rate, not the unknown channel capacity. At $r=1$ the fiber rate approaches $0.97470$ qubits per mode as $p\to\infty$ and is $0.97463$ at $p=10^4$.

High-loss free-space links are unsuitable settings for the one-way quantum communication studied here. Our Haar model requires collection of more than half of the output modes, $R>N/2$. For a physical link, one must also check the transmission eigenvalues: far-field diffraction between finite apertures can leave even the best eigenmode below $50\%$ transmission~\cite[Appendix~A.2]{Pirandola2021FreeSpace}. If every eigenmode has transmissivity at most $\tfrac12$ in every fading realization, the thermal-loss channel is antidegradable and its one-way quantum capacity vanishes at every temperature (\cref{sec:zerotemp}). Average loss alone does not establish this condition. We therefore use a low-loss microwave setting to illustrate thermal effects. Two-way assisted quantum communication (\cref{sec:twoway}) may still be possible in this high-loss regime. Low-loss free-space links are also outside this exclusion.

\section{Relation to previous work}
\label{sec:related}

Random channels whose realization is known to the receiver but not to the sender appear in the theory of measurement-induced phase transitions~\cite{LiChenFisher2018,SkinnerRuhmanNahum2019,ChanNandkishorePretkoSmith2019}. There the random channel is a circuit of random unitaries interspersed with measurements, the decoder holds the circuit and the measurement record, and the encoder knows only their distribution. Refs.~\cite{ChoiBaoQiAltman2020,GullansHuse2020} define the capacity of such a random channel by maximizing the coherent information after averaging over realizations, and relate the entanglement transition to the vanishing of this channel-averaged capacity density~\cite{BaoChoiAltman2020}. Our fading capacity is the energy-constrained bosonic version of this quantity. The receiver's copy of $U$ plays the role of the measurement record, and the Haar average replaces the circuit average. The isotropic thermal input replaces the maximally mixed input, which symmetry singles out there as well~\cite{ChoiBaoQiAltman2020}. The threshold $R=N/2$ also has a counterpart. For $T=N$ the block $K_U$ has $R$ unit and $N-R$ vanishing transmissivities for every $U$, so the channel erases $N-R$ modes in a Haar-random basis known to the receiver. \Cref{thm:capacity} then gives $Q^F_P=(2R-N)\,g(P/N)$ for $R\ge N/2$ and zero otherwise. This is the bosonic form of the $(1-2f)_+$ qubits per qubit of the quantum erasure channel at erasure probability $f=1-R/N$~\cite{BennettDiVincenzoSmolin1997}, which a random unitary encoding attains~\cite{ChoiBaoQiAltman2020}. Our discarded modes are losses of this kind, since their content stays with the environment, whereas recorded measurement outcomes raise the rate to $1-f$~\cite{ChoiBaoQiAltman2020}.

Our results also rest on known facts about the one-mode attenuator. At $\nenv=0$ the one-mode attenuator of transmissivity $\eta$ is degradable for $\eta\ge\frac12$ and antidegradable, with zero quantum capacity, for $\eta\le\frac12$~\cite{CarusoGiovannetti2006,WolfPerezGarciaGiedke2007}. Degradability makes its quantum capacity single-letter~\cite{DevetakShor2005}, and a thermal state attains this capacity under a mean photon-number constraint~\cite{WildeHaydenGuha2012,WildeQi2018}. Related results on fading bosonic channels calculate different types of capacities. When the receiver knows each realization of this channel, its two-way assisted capacities equal the fixed-transmissivity values~\cite{PirandolaEtAl2017} averaged over the fading~\cite{Pirandola2021FreeSpace}. Under a mean photon-number constraint, Ref.~\cite{PirandolaSpedalieri2026Multimode} determined the two-way capacities of a fixed passive network and of a one-mode channel whose random transmissivity is revealed to the receiver. The former depend only on the squared singular values of the transmission matrix. Ref.~\cite{MishraWilde2026Gaussian} also contains the fixed-network formula and bounds the same capacities for general multimode Gaussian channels. Without channel-state information, numerics find regimes where an optimized non-Gaussian input has positive coherent information while every thermal input under the same energy constraint has nonpositive coherent information~\cite{CatalanoEtAl2026}. A uniformly random phase rotation (our channel at $N=T=R=1$) has zero quantum capacity when neither party learns it~\cite{LamiWilde2023}. Ergodic classical and private capacities were bounded for a turbulent free-space link with about 200 spatial modes~\cite{ChandrasekaranShapiroWang2014}. 

For $T\le R$ and $T+R\le N$, the eigenvalues of $K_U^\dagger K_U$ form the Jacobi unitary ensemble with weight $x^{R-T}(1-x)^{N-R-T}$ (Theorem~2.2 of Ref.~\cite{Collins2005} and Ref.~\cite{SimonMoustakas2006}). Hence
\begin{equation}
\bar I_N(P,\nenv)=\int_0^1I_{P/T,\nenv}(x)\,\mathcal K_N(x,x)\,\dd x,
\label{eq:jacobi}
\end{equation}
where $\mathcal K_N$ is the weighted Christoffel--Darboux kernel of the first $T$ orthonormal Jacobi polynomials for this weight. This is the analogue of Telatar's Laguerre-polynomial formula (\cref{tab:telatar}). With $\log(1+\tfrac{P}{T}x)$ in place of $I_{P/T,\nenv}$, the same expression is the ergodic capacity of the classical Jacobi MIMO channel at signal-to-noise ratio $P/T$ per input mode (Theorem~1 of Ref.~\cite{DarFederShtaif2013}). With the one-mode Holevo capacity in place of $I_{P/T,\nenv}$, it is the Haar-averaged Holevo capacity of Ref.~\cite{PoplawskaJarzyna2026}. The other rank regimes reduce to this one through the exact atoms, the exchange of $T$ and $R$, and the reflection $x\mapsto1-x$ (see the remark following Theorem~2.2 of Ref.~\cite{Collins2005}). For $p>0$ and bounded $T_N-tN$, $R_N-rN$, and $P_N-pN$, a bounded remainder $\bar I_N(P_N,\nenv)=N\bar I_{t,r}(p,\nenv)+O(1)$ in \cref{thm:limit} is a separate random-matrix statement. The same kernel is the natural tool for such a bound. We do not pursue it here.

\section{Conclusion}
\label{sec:conclusion}

Telatar's formula showed that the capacity of a multiple-antenna fading channel grows linearly with the number of antennas even though the sender knows only the statistics of the channel~\cite{Telatar1999}. This result underlies the multiple-antenna links of today's cellular networks.
We have extended this derivation to quantum information sent through a Haar-random multimode network whose realization is known to the receiver but not to the sender. Under a mean photon-number constraint, the zero-temperature quantum capacity grows linearly with the number of modes whenever the receiver collects more than half of the outputs. At every temperature, the Haar-averaged coherent information of the isotropic thermal input is an achievable rate whose value per mode converges to an integral against Wachter's law. Thus linear scaling survives random mixing under the stated assumptions. 

With intrinsic loss, the same conclusion follows wherever the limiting achievable rate is positive; exactness of that rate is not established. This conclusion bears on the settings of \cref{sec:intro}: optical fibers for space-division multiplexing whose modes mix at random, multimode fibers carrying quantum light, and cryogenic microwave networks between superconducting processors. The examples in \cref{sec:numerics} use representative loss and noise scales for these settings. High-loss free-space links are excluded from unassisted quantum communication under the following condition. When every transmission eigenvalue is at most $\tfrac12$ in every realization, their one-way capacity vanishes (\cref{sec:zerotemp}), while their two-way assisted capacities may remain positive. 

The exact capacity at nonzero temperature remains open, as it does for a single thermal attenuator, and the two-way assisted capacities are left to future work (\cref{sec:twoway}). We anticipate that this work will spur a new direction in quantum information theory aimed at calculating and determining use cases for average channel capacities.

\appendices
\crefalias{section}{appendix} 

\section{The one-mode thermal attenuator}
\label{app:onemode}

This appendix fixes conventions and treats one mode. Its output is the function $I_{n,\nenv}$ of \cref{eq:Fdef}, the coherent information of a one-mode thermal attenuator with a thermal input. \Cref{app:channel} sums this function over the modes of the multimode channel. \Cref{tab:symbols} collects the notation of the appendices.

\begin{table}[!t]
\centering
\caption{\label{tab:symbols}Notation of the appendices, grouped by role. Logarithms are base two. Barred coherent-information symbols denote Haar averages or their limits. Proof-local notation is defined where used.}
\renewcommand{\arraystretch}{1.15}
\begin{tabular}{l|>{\raggedright\arraybackslash}p{0.54\textwidth}|l}
\hline
Symbol & Meaning & Location \\
\hline\hline
\multicolumn{3}{l}{\textit{Modes and photon numbers}} \\
\hline
$N$, $T$, $R$ & number of modes, of sender inputs, and of receiver outputs & \cref{app:channel} \\
$t$, $r$ & limits of $T/N$ and $R/N$ & \cref{app:channel} \\
$P$, $p$, $n$ & photon-number budget $P$ per use; $p=P/N$ per network mode; one-mode input occupation $n$, set to $P/T$ for the isotropic input & \cref{def:code}, \cref{app:channel} \\
$\nenv$ & mean photon number of each environment mode & \cref{app:onemode} \\
\hline
\multicolumn{3}{l}{\textit{States and operators}} \\
\hline
$\hat a_i$, $\hat N$, $\meanN{\rho}$ & mode operators, total number operator, and mean photon number of $\rho$ & \cref{app:onemode} \\
$g(x)$, $\tau(n)$ & entropy of a thermal mode with mean photon number $x$, and the thermal state with mean photon number $n$ & \cref{app:onemode} \\
$\isoinput{P}$ & isotropic input $\tau(P/T)^{\otimes T}$ of total photon number $P$ & \cref{app:channel} \\
$\Ucal_U$, $\chi_\rho$ & passive unitary of $U$, and characteristic function of $\rho$ & \cref{app:channel} \\
\hline
\multicolumn{3}{l}{\textit{Channels}} \\
\hline
$K_U$, $K^E_U$ & top-left $R\times T$ and bottom-left $(N-R)\times T$ blocks of $U$ & \cref{app:channel}, \cref{lem:charfun} \\
$\eta_i(U)$ & transmissivities, the eigenvalues of $K_U^\dagger K_U$ & \cref{app:channel} \\
$\Phi_{K,\nenv}$, $\Phi_U$ & thermal attenuator with block $K$, and $\Phi_{K_U,\nenv}$ & \cref{def:channel}, \cref{app:channel} \\
$\mc N_{K,\nenv}$ & classical Gaussian noise on the output & \cref{lem:charfun} \\
$\Phi_{\mathcal P}$, $\Phi_k$ & flagged channel for the partition $\mathcal P=\{\Omega_k\}$ with cell weights $w_k$, and the channel averaged over cell $k$ & \cref{app:achievability} \\
\hline
\multicolumn{3}{l}{\textit{Coherent information}} \\
\hline
$\Ic(\rho,\Phi)$ & coherent information of the state $\rho$ through the channel $\Phi$ & \cref{eq:Ic} \\
$I_{n,\nenv}(\eta)$ & coherent information of the one-mode attenuator of transmissivity $\eta$ with input $\tau(n)$ & \cref{eq:Fdef} \\
$\bar I_N(P,\nenv)$ & Haar-averaged coherent information of $\isoinput{P}$, in qubits per use; $T,R$ suppressed & \cref{eq:JN} \\
$\bar I_{t,r}(p,\nenv)$ & limit of $\bar I_N(P_N,\nenv)/N$ with $P_N/N\to p$, in qubits per network mode & \cref{eq:q} \\
\hline
\multicolumn{3}{l}{\textit{Codes and capacities}} \\
\hline
$\mc E$, $\mc D_{U^m}$, $m$, $d$ & encoder, decoders indexed by the realizations, block length, and code dimension & \cref{def:code} \\
$Q^F_P$ & fading capacity with mean photon-number budget $P$ per use & \cref{def:code} \\
$Q_P(\Phi_{\mathcal P})$ & energy-constrained entanglement-transmission capacity of the flagged channel & \cref{app:achievability} \\
\hline
\multicolumn{3}{l}{\textit{Transmissivity laws}} \\
\hline
$\nu_N$, $\nu_{t,r}$ & expected empirical transmissivity measure and its Wachter limit, of masses $T/N$ and $t$ & \cref{app:limit} \\
$w_{t,r}$, $\alpha_\pm$ & density of the Wachter bulk and its endpoints & \cref{eq:wachter} \\
$\mathcal K_N$ & weighted Christoffel--Darboux kernel of the Jacobi ensemble & \cref{eq:jacobi} \\
\hline
\end{tabular}
\end{table}

\textit{Conventions.} Logarithms are base two, and
\begin{equation}
g(x)=(x+1)\log(x+1)-x\log x
\end{equation}
is the entropy of a thermal mode with mean photon number $x$. Let $\hat a_1,\dots,\hat a_M$ be the mode operators of $M$ modes and $\hat N=\sum_i\hat a_i^\dagger\hat a_i$ the total number operator. A state of $M$ modes is a density operator $\rho$ on the Fock space of $M$ modes with finite mean photon number $\meanN\rho=\Tr\rho\hat N$. All entropies below are then finite, and $H(X|Y)=H(XY)-H(Y)$. The one-mode thermal state $\tau(n)$ is the Gibbs state of $\hat a^\dagger\hat a$ with $\meanN{\tau(n)}=n$, and $H(\tau(n))=g(n)$.

\textit{The one-mode thermal attenuator.} Let $\hat a$ be the input mode and $\hat e$ an environment mode prepared in $\tau(\nenv)$ with $\nenv\ge0$. The beam splitter of transmissivity $\eta\in[0,1]$ is the unitary $\mc B_\eta$ on the two modes with $\mc B_\eta^\dagger\hat a\,\mc B_\eta=\sqrt\eta\,\hat a+\sqrt{1-\eta}\,\hat e$ and $\mc B_\eta^\dagger\hat e\,\mc B_\eta=\sqrt{1-\eta}\,\hat a-\sqrt\eta\,\hat e$. The thermal attenuator $\Phi_{\sqrt\eta,\nenv}$ maps a state $\rho$ of the input mode to the reduced state of the mode $\hat a$ in $\mc B_\eta\parens[\big]{\rho\otimes\tau(\nenv)}\mc B_\eta^\dagger$. The notation anticipates \cref{def:channel}, where this channel is the case of the $1\times1$ matrix $K=(\sqrt\eta)$. For $\eta=1$ the attenuator is the identity, and for $\eta=0$ it replaces every input by $\tau(\nenv)$.

\textit{Coherent information.} For a state $\rho$ and a channel $\Phi$, let $\psi$ be a purification of $\rho$ on a reference system $A$, and set
\begin{equation}
\Ic(\rho,\Phi)=H(B)-H(AB),
\label{eq:Ic}
\end{equation}
evaluated on $(\id_A\otimes\Phi)(\psi)$. The value does not depend on the purification. Three properties follow directly from \cref{eq:Ic}. A unitary channel $\mc V$ after $\Phi$ leaves $\Ic$ unchanged, and one before $\Phi$ acts on the input: $\Ic(\rho,\Phi\circ\mc V)=\Ic(\mc V(\rho),\Phi)$. Coherent information is additive on product inputs to product channels, because purifications and entropies of products add. If $\Phi$ is realized by an isometry with a pure environment $E$, then $H(AB)=H(E)$ and $\Ic(\rho,\Phi)=H(B)-H(E)$.

\textit{The one-mode function.} We set
\begin{equation}
I_{n,\nenv}(\eta)=\Ic\parens[\big]{\tau(n),\Phi_{\sqrt\eta,\nenv}},\qquad n,\nenv\ge0,\ \eta\in[0,1].
\label{eq:Fdef}
\end{equation}
This is the coherent information of the one-mode thermal attenuator with a thermal input, evaluated by Holevo and Werner~\cite{HolevoWerner2001}. At $\nenv=0$ and $\eta\ge\frac12$ it is also the quantum capacity of the attenuator under the mean photon-number constraint $n$~\cite{WildeHaydenGuha2012,WildeQi2018}.

\begin{lemma}[One-mode coherent information]
\label{lem:onemode}
For $n,\nenv\ge0$ and $\eta\in[0,1]$,
\begin{align}
I_{n,\nenv}(\eta)&=g\parens[\big]{\eta n+(1-\eta)\nenv}-g\parens[\big]{u_+(\eta)}-g\parens[\big]{u_-(\eta)},\label{eq:F}\\
u_\pm(\eta)&=\tfrac12\squ[\big]{d_\eta-1\pm(1-\eta)(n-\nenv)},\notag\\
d_\eta&=\sqrt{\squ[\big]{(1+\eta)n+(1-\eta)\nenv+1}^2-4\eta n(n+1)}.\notag
\end{align}
\end{lemma}

\Cref{eq:F} is stated as (7) of Ref.~\cite{NohPirandolaJiang2020}, whose $D$ is our $d_\eta$. Note that $u_\pm\ge0$, because $d_\eta^2-[1+(1-\eta)(n-\nenv)]^2=4(1-\eta)(1+n)\nenv$ and $d_\eta^2-[1-(1-\eta)(n-\nenv)]^2=4(1-\eta)(1+\nenv)n$.


We record the properties of $I_{n,\nenv}$ used below. Three follow from the definition~\eqref{eq:Fdef} alone. First, $I_{n,\nenv}(0)=-g(n)$, because $\Phi_{0,\nenv}$ replaces the input by $\tau(\nenv)$ while the reference keeps entropy $g(n)$. Second, $I_{n,\nenv}(1)=g(n)$, because $\Phi_{1,\nenv}$ is the identity. Third, $I_{0,\nenv}\equiv0$. Two more follow from \cref{eq:F}. The function $I$ is jointly continuous in $(n,\eta)$ on $[0,\infty)\times[0,1]$. At $\nenv=0$ one has $d_\eta=(1-\eta)n+1$, so
\begin{equation}
I_{n,0}(\eta)=g(\eta n)-g\parens[\big]{(1-\eta)n},
\label{eq:Fzero}
\end{equation}
which is antisymmetric about $\eta=\frac12$.

\section{The multimode channel and its fading capacity}
\label{app:channel}

This appendix defines the multimode thermal attenuator, the random model, and the fading capacity. Its main output is \cref{thm:singleuse}: the coherent information of the isotropic thermal input is the sum of the one-mode function $I_{n,\nenv}$ over the transmissivities of the realization. \Cref{thm:capacity} then states what the Haar average of this sum says about the fading capacity, with proofs in \cref{app:achievability,app:exact}, and the last paragraph sets up the large-$N$ limit.

\textit{Passive unitaries and characteristic functions.} We write $U(M)$ for the unitary group with Haar probability measure $\dd U$, so that $\int\dd U\,f(U)$ is the Haar average of a function $f$. For $U\in U(M)$ let $\Ucal_U$ be the passive unitary on $M$ modes with $\Ucal_U^\dagger\hat a_i\Ucal_U=\sum_jU_{ij}\hat a_j$. The map $U\mapsto\Ucal_U$ is a strongly continuous unitary representation of $U(M)$~\cite{Holevo2019,Serafini2017}, and the beam splitter $\mc B_\eta$ of \cref{app:onemode} is $\Ucal_U$ for $U=\bigl(\begin{smallmatrix}\sqrt\eta&\sqrt{1-\eta}\\\sqrt{1-\eta}&-\sqrt\eta\end{smallmatrix}\bigr)$. The convenient description of passive unitaries is through characteristic functions. For $z\in\C^M$ let $D(z)=\exp\sum_i(z_i\hat a_i^\dagger-\bar z_i\hat a_i)$ and $\chi_\rho(z)=\Tr\rho D(z)$. A state is determined by its characteristic function. The characteristic function of a product state is the product of the factors' characteristic functions,
\begin{equation}
\chi_{\tau(\nenv)}(w)=\exp\squ[\big]{-(\nenv+\tfrac12)\abs{w}^2},
\end{equation}
and
\begin{equation}
\chi_{\Ucal_U\rho\,\Ucal_U^\dagger}(z)=\chi_\rho(U^\dagger z).
\label{eq:passive}
\end{equation}

\begin{definition}[Thermal attenuator]
\label{def:channel}
Let $K$ be an $R\times T$ complex matrix with $\norm{K}\le1$ and let $\nenv\ge0$. Choose $N$ and a unitary $U\in U(N)$ whose top-left $R\times T$ block is $K$. The thermal attenuator $\Phi_{K,\nenv}$ maps a state $\rho$ of $T$ modes to the state of the first $R$ output modes of $\Ucal_U\parens[\big]{\rho\otimes\tau(\nenv)^{\otimes(N-T)}}\Ucal_U^\dagger$. The cases $T=0$ and $R=0$ are allowed. The one-mode attenuator $\Phi_{\sqrt\eta,\nenv}$ of \cref{app:onemode} is the case $N=2$, $T=R=1$, and $K=(\sqrt\eta)$.
\end{definition}

A unitary $U$ as in the definition exists for $N=T+R$, for example the block matrix with rows $(K,\sqrt{I-KK^\dagger})$ and $(\sqrt{I-K^\dagger K},-K^\dagger)$. In the language of multimode bosonic Gaussian channels~\cite{EisertWolf2007,CarusoEisertGiovannettiHolevo2008,WeedbrookEtAl2012}, $\Phi_{K,\nenv}$ is a Gaussian channel with a passive unitary dilation whose environment consists of $N-T$ thermal modes. Ref.~\cite{CarusoEisertGiovannettiHolevo2011} determines the minimal number of environment modes in such dilations for every Gaussian channel. The following lemma shows that $\Phi_{K,\nenv}$ does not depend on the choice of $N$ and $U$. It also supplies the composition rule behind the factorization of \cref{lem:factor} and the environment identities of \cref{app:exact}.

\begin{lemma}
\label{lem:charfun}
For every state $\rho$ of $T$ modes and every $z\in\C^R$,
\begin{equation}
\chi_{\Phi_{K,\nenv}(\rho)}(z)=\chi_\rho(K^\dagger z)\,\exp\squ[\big]{-(\nenv+\tfrac12)\ang{z,(I-KK^\dagger)z}}.
\label{eq:charfun}
\end{equation}
In particular, $\Phi_{K,\nenv}$ depends only on $K$ and $\nenv$. Moreover:
\begin{enumerate}
\item[(a)] $\Phi_{V,0}(\rho)=\Ucal_V\rho\,\Ucal_V^\dagger$ for unitary $V$, and $\Phi_{WKV,\nenv}=\Phi_{W,0}\circ\Phi_{K,\nenv}\circ\Phi_{V,0}$ for $W\in U(R)$ and $V\in U(T)$.
\item[(b)] $\Phi_{K,\nenv}=\mc N_{K,\nenv}\circ\Phi_{K,0}$, where $\mc N_{K,\nenv}$ is the channel that multiplies $\chi$ by $\exp[-\nenv\ang{z,(I-KK^\dagger)z}]$ (classical Gaussian noise).
\item[(c)] Discarding all but the first $R'$ output modes gives $\Phi_{K',\nenv}$, where $K'$ consists of the first $R'$ rows of $K$.
\item[(d)] For $T=0$ the output is $\tau(\nenv)^{\otimes R}$, and for $R=0$ the channel is the trace.
\item[(e)] For $\nenv=0$, the joint state of a purifying reference $A$, the first $R$ output modes $B$, and the remaining $N-R$ output modes $E$ is pure. The state of $E$ is then $\Phi_{K^E,0}(\rho)$, with $K^E$ the bottom-left $(N-R)\times T$ block of $U$.
\end{enumerate}
\end{lemma}

\begin{proof}
Let $U$ have top-left block $K$, and let $L$ be its top-right $R\times(N-T)$ block. Put $\sigma=\rho\otimes\tau(\nenv)^{\otimes(N-T)}$. By \cref{eq:passive}, $\chi_{\Ucal_U\sigma\Ucal_U^\dagger}(z)=\chi_\sigma(U^\dagger z)$. Tracing out the last $N-R$ output modes restricts $\chi$ to vectors $z$ supported on the first $R$ coordinates. For such $z$, $U^\dagger z=(K^\dagger z,L^\dagger z)$, so
\begin{equation}
\chi_\sigma(U^\dagger z)=\chi_\rho(K^\dagger z)\,\exp\squ[\big]{-(\nenv+\tfrac12)\norm{L^\dagger z}^2}.
\end{equation}
The first $R$ rows of $U$ are orthonormal, so $KK^\dagger+LL^\dagger=I$, which gives \cref{eq:charfun}. The right side involves only $K$ and $\nenv$.

Items (a) through (d) are substitutions into \cref{eq:charfun}. For (a), $I-VV^\dagger=0$ and $\chi_\rho(V^\dagger z)$ is the characteristic function of $\Ucal_V\rho\Ucal_V^\dagger$. The composition rule follows from $\chi_{\Phi_{V,0}(\rho)}(K^\dagger z)=\chi_\rho(V^\dagger K^\dagger z)$ and $\ang{W^\dagger z,(I-KK^\dagger)W^\dagger z}=\ang{z,(I-WKK^\dagger W^\dagger)z}$. For (b), split the exponent as $\frac12+\nenv$. For (c), set the discarded coordinates of $z$ to zero. For (d), note that $K$ is empty. For (e), the environment input is the vacuum, so $\Ucal_U$ acts on a purification of $\rho\otimes\ketbra{0}^{\otimes(N-T)}$ and the joint state of $A$, $B$, and $E$ is pure. The same computation with $z$ supported on the last $N-R$ coordinates identifies the state of $E$ as $\Phi_{K^E,0}(\rho)$.
\end{proof}

\begin{lemma}[Factorization]
\label{lem:factor}
Let $K=W\Sigma V^\dagger$ be a singular value decomposition with $W\in U(R)$, $V\in U(T)$, and $\Sigma_{ii}=\sqrt{\eta_i}$ for $i\le\ell=\min\{T,R\}$. Then
\begin{equation}
\Phi_{K,\nenv}=\Phi_{W,0}\circ\parens[\Big]{\bigotimes_{i=1}^{\ell}\Phi_{\sqrt{\eta_i},\nenv}\otimes\mc T^{\otimes(T-\ell)}\otimes\mc P_{\nenv}^{\otimes(R-\ell)}}\circ\Phi_{V^\dagger,0},
\label{eq:factor}
\end{equation}
where $\mc T$ is the trace of one input mode and $\mc P_\nenv$ prepares one output mode in $\tau(\nenv)$. At most one of the last two factors is present.
\end{lemma}

\begin{proof}
By \cref{lem:charfun}(a), $\Phi_{K,\nenv}=\Phi_{W,0}\circ\Phi_{\Sigma,\nenv}\circ\Phi_{V^\dagger,0}$. By \cref{eq:charfun},
\begin{align}
\chi_{\Phi_{\Sigma,\nenv}(\rho)}(z)&=\chi_\rho(\Sigma^\dagger z)\prod_{i\le\ell}e^{-(\nenv+\frac12)(1-\eta_i)\abs{z_i}^2}\notag\\
&\quad\times\prod_{\ell<i\le R}e^{-(\nenv+\frac12)\abs{z_i}^2},
\end{align}
and $\Sigma^\dagger z$ vanishes in the input coordinates $i>\ell$. Read mode by mode, this is the characteristic function of the product channel in \cref{eq:factor}: coordinate $i\le\ell$ carries $\Phi_{\sqrt{\eta_i},\nenv}$, an input coordinate $i>\ell$ is traced out, and an output coordinate $i>\ell$ is prepared in $\tau(\nenv)$.
\end{proof}

\textit{The random model.} Fix integers $N\ge1$, $1\le T\le N$, $0\le R\le N$, and $\nenv\ge0$. For $U\in U(N)$ let $K_U$ be its top-left $R\times T$ block, and let $\eta_1(U),\dots,\eta_T(U)\in[0,1]$ be the eigenvalues of $K_U^\dagger K_U$, the transmissivities. The random channel is
\begin{equation}
\Phi_U=\Phi_{K_U,\nenv}
\end{equation}
with $U$ Haar distributed. Write $\isoinput{P}:=\tau(P/T)^{\otimes T}$ for the isotropic input of photon number $P$; it is a function of $\hat N$. The case $T=0$ has zero capacity and is excluded from formulas involving $P/T$.

\begin{definition}[Fading code and fading capacity]
\label{def:code}
A fading code of block length $m$ and dimension $d$ consists of an encoder $\mc E$ from operators on $\C^d$ to states of $mT$ modes and a measurable family of decoders $\mc D_{U^m}$. Each decoder, indexed by $U^m=(U_1,\dots,U_m)\in U(N)^m$, maps states of $mR$ modes to operators on $\hat B\cong\C^d$. Let $\Fme_{CC'}$ be the maximally entangled state of the encoder's input $C'\cong\C^d$ with a reference $C$, and let $\pi_d$ be the maximally mixed state. (The letter $R$ is reserved for the number of receiver modes.) The photon number of the code is $\meanN{\mc E(\pi_d)}$, and its error is
\begin{equation}
1-\int\dd U^m\,F\parens[\Big]{\Fme_{C\hat B},\,(\id_C\otimes\mc D_{U^m}\circ(\Phi_{U_1}\otimes\cdots\otimes\Phi_{U_m})\circ\mc E)(\Fme_{CC'})},
\end{equation}
with $\dd U^m=\dd U_1\cdots\dd U_m$ the product Haar measure and $F(\Fme_{C\hat B},\omega)=\Tr\Fme_{C\hat B}\,\omega$ the fidelity with the pure state $\Fme_{C\hat B}$. A rate $\mc R$ is achievable with photon number $P$ if, for every $\delta>0$, codes exist with $\log d\ge m(\mc R-\delta)$, photon number at most $mP$, and error tending to zero. The fading capacity $Q^F_P$ is the supremum of the achievable rates.
\end{definition}

\Cref{def:code} is entanglement transmission with an average energy constraint, the setting of the achievability theorem of Wilde and Qi~\cite{WildeQi2018}. Quantum communication under a uniform energy constraint has the same capacity. For a fixed channel this is Theorem~1 of Ref.~\cite{WildeQi2018}, and its proof applies to the channel from $C'$ to $\hat B$ averaged over $U^m$. Hence nothing below depends on this choice. We use the present definition because its error is linear in the decoded state. The isotropic input $\isoinput{P}$ has photon number exactly $P$ and therefore saturates the constraint.

\begin{theorem}[Single-use coherent information]
\label{thm:singleuse}
For every $\nenv\ge0$, $U\in U(N)$, and $P\ge0$,
\begin{equation}
\Ic\parens[\big]{\isoinput{P},\Phi_U}=\sum_{i=1}^{T}I_{P/T,\nenv}\parens[\big]{\eta_i(U)}.
\label{eq:singleuse}
\end{equation}
\end{theorem}

\begin{proof}
Set $n=P/T$. Apply \cref{lem:factor} to $K_U$. The isotropic state $\isoinput{P}$ is a function of $\hat N$, so it is fixed by $\Phi_{V^\dagger,0}$, and the unitary channel $\Phi_{W,0}$ on the output does not change $\Ic$. Therefore
\begin{equation}
\Ic\parens[\big]{\isoinput{P},\Phi_U}=\Ic\parens[\big]{\isoinput{P},\Phi_{\Sigma,\nenv}}.
\end{equation}
The right side is a product input to a product channel, so it is the sum of the single-mode values. These are $\Ic(\tau(n),\Phi_{\sqrt{\eta_i},\nenv})=I_{n,\nenv}(\eta_i)$ for $i\le\ell$, $\Ic(\tau(n),\mc T)=0-g(n)=I_{n,\nenv}(0)$ for each traced-out input mode, and $\Ic(1,\mc P_\nenv)=g(\nenv)-g(\nenv)=0$ for each prepared output mode. When $T>R$, the eigenvalues of $K_U^\dagger K_U$ are $\eta_1,\dots,\eta_\ell$ together with $T-\ell$ zeros, one for each traced-out direction. Thus the sum runs over all $T$ eigenvalues in every case, which is \cref{eq:singleuse}.
\end{proof}

In other words, the multimode attenuator with the isotropic input behaves as $T$ independent one-mode attenuators with transmissivities $\eta_i(U)$, and no multimode entropy is ever computed. The Haar average of \cref{eq:singleuse} at the isotropic input of photon number $P$ is
\begin{equation}
\bar I_N(P,\nenv)=\int\dd U\sum_{i=1}^{T}I_{P/T,\nenv}\parens[\big]{\eta_i(U)},
\label{eq:JN}
\end{equation}
the averaged coherent information of the isotropic input. The bar marks the Haar average, and the subscript $N$ the size of the network. The next theorem relates it to the fading capacity.

\begin{theorem}[Fading capacity]
\label{thm:capacity}
Let $\nenv\ge0$.
\begin{enumerate}
\item[(i)] For all $N$, $T$, $R$, and $P$, $Q^F_P\ge \bar I_N(P,\nenv)$.
\item[(ii)] At $\nenv=0$, for all $N$, $T$, and $P$,
\begin{equation}
Q^F_P=\begin{cases}\bar I_N(P,0), & R\ge N/2,\\ 0, & R\le N/2.\end{cases}
\end{equation}
\end{enumerate}
\end{theorem}

Part (i) is proved in \cref{app:achievability} and part (ii) in \cref{app:exact}.

\textit{The large-$N$ limit.} We now let $N\to\infty$ with $T_N/N\to t\in(0,1]$, $R_N/N\to r\in[0,1]$, and $P_N/N\to p\ge0$, so that the photon number per mode is fixed. By \cref{eq:JN}, the rate per mode $\bar I_N(P_N,\nenv)/N$ is the integral of the one-mode function $I_{P_N/T_N,\nenv}$ against the expected empirical measure of the transmissivities. Let $\Pi_T$ and $\Pi_R$ be the coordinate projections of ranks $T$ and $R$ in $\C^N$. The matrix $K_U^\dagger K_U$ is $\Pi_TU^\dagger\Pi_RU\Pi_T$ restricted to the first $T$ coordinates, the compression of one coordinate projection by a Haar-rotated one. Its eigenvalues follow a Jacobi ensemble~\cite{Collins2005,SimonMoustakas2006}, and their empirical measure converges to Wachter's law~\cite{Wachter1980,Collins2005}. \Cref{app:limit} carries out this limit.

\section{The large-\texorpdfstring{$N$}{N} limit}
\label{app:limit}

This appendix proves the main result, \cref{thm:limit}: under the scaling of the last paragraph of \cref{app:channel}, the averaged coherent information per mode converges to an integral of $I_{p/t,\nenv}$ against Wachter's law. \Cref{cor:permode} translates it into a statement about the fading capacity.

\textit{Integral form.} Let
\begin{equation}
\nu_N=\frac1N\int\dd U\sum_{i=1}^{T}\delta_{\eta_i(U)}
\end{equation}
be the expected empirical measure of the transmissivities, a measure on $[0,1]$ of total mass $T/N$. By \cref{eq:JN},
\begin{equation}
\frac1N\bar I_N(P,\nenv)=\int I_{P/T,\nenv}\,\dd\nu_N.
\label{eq:integralform}
\end{equation}
The limit therefore requires the limit of $\nu_N$ and the continuity of the integrand. The same structure appears for lossy bosonic memory channels, whose capacities are integrals of one-mode capacities against a limiting transmissivity law given by Szeg\H{o}'s theorem~\cite{LupoGiovannettiMancini2010}.

\textit{Wachter measure.} For $t\in(0,1]$ and $r\in[0,1]$ let $\alpha_\pm=\parens[\big]{\sqrt{t(1-r)}\pm\sqrt{r(1-t)}}^2$ and
\begin{align}
\nu_{t,r}&=w_{t,r}(x)\,\dd x+(t-r)_+\,\delta_0+(t+r-1)_+\,\delta_1,\label{eq:wachter}\\
w_{t,r}(x)&=\frac{\sqrt{(\alpha_+-x)(x-\alpha_-)}}{2\pi x(1-x)}\,\mathbf 1_{[\alpha_-,\alpha_+]}(x),\notag
\end{align}
where $(x)_+=\max\{x,0\}$ and the density term is absent when $\alpha_-=\alpha_+$. The measure $\nu_{t,r}$ has total mass $t$. We define
\begin{align}
\bar I_{t,r}(p,\nenv)&=\int I_{p/t,\nenv}\,\dd\nu_{t,r}\notag\\
&=\int_{\alpha_-}^{\alpha_+}I_{p/t,\nenv}(x)\,w_{t,r}(x)\,\dd x+\big[(t+r-1)_+-(t-r)_+\big]\,g(p/t),
\label{eq:q}
\end{align}
where the second form uses the endpoint values $I_{n,\nenv}(0)=-g(n)$ and $I_{n,\nenv}(1)=g(n)$ of \cref{app:onemode}. The subscript $t,r$ marks the limiting shape of the network, and the argument $p$ is the photon number per mode.

\begin{lemma}[Wachter's law]
\label{lem:wachter}
If $T/N\to t\in(0,1]$ and $R/N\to r\in[0,1]$, then $\nu_N\to\nu_{t,r}$ weakly.
\end{lemma}

\begin{proof}
Recall from the end of \cref{app:channel} that $K_U^\dagger K_U$ is the compression $\Pi_TU^\dagger\Pi_RU\Pi_T$, where $\Pi_T$ and $\Pi_R$ are the coordinate projections of ranks $T$ and $R$. The spectral law of $\Pi_T$ is the Bernoulli law $(1-T/N)\delta_0+(T/N)\delta_1$, and conjugation by $U$ does not change the spectral law of $\Pi_R$. What the Haar measure fixes is the joint law of the pair. By Voiculescu's theorem, constant diagonal matrices with limiting spectral laws are asymptotically free from an independent Haar unitary as $N\to\infty$~\cite{Voiculescu1991}, and therefore $\Pi_T$ and $U^\dagger\Pi_RU$ are asymptotically free (see also Ref.~\cite{NS06}). Hence the expected normalized moments of $\Pi_TU^\dagger\Pi_RU\Pi_T$ converge to those of the free multiplicative convolution of the Bernoulli laws with weights $t$ and $r$ at $1$. This measure is given in Sec.~3.2 of Ref.~\cite{Collins2005}, see also Example~3.6.7 of Ref.~\cite{VDN1992} and Ref.~\cite{Wachter1980}, and equals
\begin{equation}
\squ[\big]{1-\min\{t,r\}}\delta_0+(t+r-1)_+\delta_1+w_{t,r}(x)\,\dd x.
\end{equation}
The matrix $\Pi_TU^\dagger\Pi_RU\Pi_T$ has the $T$ eigenvalues $\eta_i(U)$ together with $N-T$ zeros, so its expected spectral measure is $\nu_N+(1-T/N)\delta_0$. Subtracting the limit of the second term gives $\nu_{t,r}$, since $1-\min\{t,r\}-(1-t)=(t-r)_+$. All measures involved live on $[0,1]$, where moment convergence is weak convergence.
\end{proof}

The atoms can be cross-checked at finite $N$. Almost surely $K_U$ has rank $\min\{T,R\}$ and $K^E_U$ has rank $\min\{T,N-R\}$, so exactly $(T-R)_+$ of the $\eta_i$ equal $0$ and exactly $(T+R-N)_+$ equal $1$.

\begin{theorem}[Large-$N$ limit]
\label{thm:limit}
Let $\nenv\ge0$. If $T_N/N\to t\in(0,1]$, $R_N/N\to r\in[0,1]$, and $P_N/N\to p\ge0$, then
\begin{equation}
\frac1N \bar I_N(P_N,\nenv)\to \bar I_{t,r}(p,\nenv).
\end{equation}
\end{theorem}

\begin{proof}
Write $n_N=P_N/T_N$ and $n=p/t$, so $n_N\to n$. By \cref{eq:integralform}, $\frac1N\bar I_N(P_N,\nenv)=\int I_{n_N,\nenv}\,\dd\nu_N$. Since $I$ is jointly continuous on the compact set $[0,n+1]\times[0,1]$, $I_{n_N,\nenv}\to I_{n,\nenv}$ uniformly on $[0,1]$, and $\nu_N$ has mass $T_N/N\le1$. Replacing $n_N$ by $n$ therefore costs $o(1)$. Next, $I_{n,\nenv}$ is continuous on $[0,1]$ and $\nu_N\to\nu_{t,r}$ weakly by \cref{lem:wachter}, so $\int I_{n,\nenv}\,\dd\nu_N\to\int I_{n,\nenv}\,\dd\nu_{t,r}=\bar I_{t,r}(p,\nenv)$. Expanding the atoms with $I_{n,\nenv}(0)=-g(n)$ and $I_{n,\nenv}(1)=g(n)$ yields the second form in \cref{eq:q}.
\end{proof}

\begin{corollary}[Capacity per mode]
\label{cor:permode}
Under the scaling of \cref{thm:limit}, $\liminf_NQ^F_{P_N}/N\ge \bar I_{t,r}(p,\nenv)$ for every $\nenv\ge0$. At $\nenv=0$ and $r\ge\frac12$, $Q^F_{P_N}/N\to \bar I_{t,r}(p,0)$, where
\begin{align}
\bar I_{t,r}(p,0)&=\int_{\alpha_-}^{\alpha_+}\squ[\Big]{g\parens[\big]{\tfrac{p}{t}x}-g\parens[\big]{\tfrac{p}{t}(1-x)}}w_{t,r}(x)\,\dd x\notag\\
&\quad+\big[(t+r-1)_+-(t-r)_+\big]\,g(p/t).
\end{align}
\end{corollary}

\begin{proof}
The bound follows from \cref{thm:capacity}(i) and \cref{thm:limit}. The formula for $\bar I_{t,r}(p,0)$ is \cref{eq:q} with $I_{n,0}$ from \cref{eq:Fzero}. For $r\ge\frac12$ and $N$ large, $R_N\ge N/2$ eventually, except along the boundary case $r=\frac12$ with $R_N<N/2$, where $Q^F_{P_N}=0=\bar I_{t,1/2}(p,0)$ by \cref{thm:capacity}(ii), the antisymmetry of $I_{n,0}$ about $\eta=\frac12$, the symmetry $w_{t,1/2}(x)=w_{t,1/2}(1-x)$, and cancellation of the two endpoint contributions. In all other cases $Q^F_{P_N}/N=\bar I_N(P_N,0)/N\to \bar I_{t,r}(p,0)$ by \cref{thm:capacity}(ii) and \cref{thm:limit}.
\end{proof}

\section{Achievability}
\label{app:achievability}

This appendix proves \cref{thm:capacity}(i): the averaged coherent information $\bar I_N(P,\nenv)$ is an achievable rate. For finitely many realizations in finite dimension, this quantity is the coherent information of a channel with a classical flag, and its achievability is the usual coding theorem, which is how Refs.~\cite{ChoiBaoQiAltman2020,GullansHuse2020} treat random circuits. What requires proof here is the continuum of realizations and the energy constraint on an infinite-dimensional output. A finite partition of $U(N)$ with a continuity bound handles the first, and the energy-constrained theorem of Wilde and Qi~\cite{WildeQi2018} handles the second.

\begin{proof}[Proof of \cref{thm:capacity}(i)]
Fix a finite Borel partition $\mathcal P=\{\Omega_k\}_{k=1}^{c}$ of $U(N)$ with $w_k=\int_{\Omega_k}\dd U>0$, and let
\begin{equation}
\Phi_{\mathcal P}(\rho)=\sum_{k=1}^{c}w_k\,\Phi_k(\rho)\otimes\ketbra{k},\qquad\Phi_k=\frac1{w_k}\int_{\Omega_k}\Phi_U\,\dd U,
\end{equation}
a channel with output $R$ modes and a $c$-valued classical flag. A code for $\Phi_{\mathcal P}$ is a fading code at the same rate, photon number, and error, because the receiver knows $U^m$ and can compute the cell indices. Hence $Q^F_P\ge Q_P(\Phi_{\mathcal P})$, where $Q_P$ is the energy-constrained entanglement-transmission capacity of Ref.~\cite{WildeQi2018}. Its Theorem~2 applies. The Gibbs hypothesis holds because $\Tr e^{-\beta\hat N}=(1-e^{-\beta})^{-T}<\infty$ for $\beta>0$. Its Condition~1 holds because an input with photon number at most $P$ has output entropy at most $\log c+R\,g(P/R+\nenv)$ by the thermal entropy bound, and at most $\log c$ for $R=0$. Thus $Q_P(\Phi_{\mathcal P})\ge\Ic(\isoinput{P},\Phi_{\mathcal P})$, and since the flag is classical, \cref{eq:Ic} gives $\Ic(\isoinput{P},\Phi_{\mathcal P})=\sum_kw_k\,\Ic(\isoinput{P},\Phi_k)$.

It remains to compare $\Ic(\isoinput{P},\Phi_k)$ with the average of $\Ic(\isoinput{P},\Phi_U)$ over the cell. Let $\omega_U=(\id_A\otimes\Phi_U)(\psi)$ for a purification $\psi$ of $\isoinput{P}$ whose reference $A$ consists of $T$ modes in the state $\isoinput{P}$. The map $U\mapsto\omega_U$ is trace-norm continuous by strong continuity of $U\mapsto\Ucal_U$, hence uniformly continuous on the compact group. Choose the partition so that $\frac12\norm{\omega_U-\omega_{U'}}_1\le\delta$ whenever $U$ and $U'$ lie in the same cell. By convexity, $\frac12\norm{\omega_k-\omega_U}_1\le\delta$ for $U\in\Omega_k$, where $\omega_k=(\id_A\otimes\Phi_k)(\psi)$. The reference has photon number $P$, so the energy-constrained continuity bound for conditional entropy of Ref.~\cite{Winter2016} gives $\epsilon(\delta)\to0$ with
\begin{equation}
\Ic(\isoinput{P},\Phi_k)\ge\frac1{w_k}\int_{\Omega_k}\Ic(\isoinput{P},\Phi_U)\,\dd U-\epsilon(\delta).
\end{equation}
Summing over $k$ with weights $w_k$ and using \cref{thm:singleuse} gives $Q^F_P\ge\bar I_N(P,\nenv)-\epsilon(\delta)$, and $\delta\to0$ completes the proof. Only the passage from codes for $\Phi_{\mathcal P}$ to fading codes was used, and only for finite flags.
\end{proof}

\section{Exactness at zero temperature}
\label{app:exact}

This appendix proves \cref{thm:capacity}(ii). Throughout, $\nenv=0$, so $\Phi_U=\Phi_{K_U,0}$. The argument for $R\ge N/2$ has three parts. A pointwise identity (\cref{lem:degrade}) shows that the receiver can produce the environment of a different realization by discarding modes. Its Haar average (\cref{lem:averaged}) makes the averaged coherent information nonnegative and expresses it as a conditional entropy. One relative-entropy inequality (\cref{lem:tangent}) then bounds the averaged coherent information of any input by an affine function of its photon number, which proves that the isotropic input is optimal.

For $R\ge N/2$ split the receiver as $B=B_1B_2$, where $B_1$ consists of the first $N-R$ output modes and $B_2$ of the remaining $2R-N$. Let $S$ be the permutation matrix that exchanges output modes $1,\dots,N-R$ with $R+1,\dots,N$, and recall from \cref{lem:charfun}(e) that $K^E_U$ is the bottom-left $(N-R)\times T$ block of $U$.

\begin{lemma}[Pointwise degradation]
\label{lem:degrade}
Let $R\ge N/2$. For every $U\in U(N)$,
\begin{equation}
\Tr_{B_2}\circ\,\Phi_U=\Phi_{K^E_{SU},0},
\label{eq:degrade}
\end{equation}
and $\Phi_{K^E_U,0}$ is the environment output of $\Phi_U$ in the sense of \cref{lem:charfun}(e), so that $\Ic(\rho,\Phi_U)=H(B)-H(E)$.
\end{lemma}

\begin{proof}
Both sides of \cref{eq:degrade} equal $\Phi_{K',0}$ with $K'$ the first $N-R$ rows of $U$ restricted to the first $T$ columns. For the left side this is \cref{lem:charfun}(c). For the right side, the last $N-R$ rows of $SU$ are the first $N-R$ rows of $U$. The second claim is \cref{lem:charfun}(e).
\end{proof}

In other words, the receiver can produce the environment's state of the realization $SU$ by discarding $B_2$. Since $U\mapsto SU$ preserves the Haar measure, this is a degradation on average, even though a single realization need not be degradable. In the language of Ref.~\cite{KianvashFanizzaGiovannetti2022}, the receiver's knowledge of $U$ is a classical flag on the mixture of the channels $\Phi_U$. Discarding $B_2$ and relabeling the flag by $U\mapsto SU$ degrades this flagged channel. We do not invoke the single-letter formula for degradable channels~\cite{DevetakShor2005}, or its energy-constrained form~\cite{WildeQi2018}, on this flagged channel. Both are stated for outputs of finite entropy, which a continuous classical flag does not have. A finite coarse-graining of the flag need not preserve degradability: its complementary output can retain information about the unresolved realization. The exact-flag degradation therefore does not by itself supply a degrading map for the finite-flag channel. \Cref{lem:tangent} below works on the channels $\Phi_U$ directly. Its bound is linear in the number of uses, which single-letterizes the converse, and it also identifies the optimal input.

\begin{lemma}[Averaged identity]
\label{lem:averaged}
Let $R\ge N/2$ and let $\rho^{(m)}$ be any state of $mT$ modes. With $\omega=(\Phi_{U_1}\otimes\cdots\otimes\Phi_{U_m})(\rho^{(m)})$,
\begin{equation}
\int\dd U^m\,\Ic\parens[\Big]{\rho^{(m)},\bigotimes_{i=1}^m\Phi_{U_i}}=\int\dd U^m\,H(B_2^m|B_1^m)_\omega\ \ge\ 0.
\label{eq:averaged}
\end{equation}
\end{lemma}

\begin{proof}
By \cref{lem:charfun}(e) applied to each factor, the product channel has a pure environment $E^m$ whose state is $(\Phi_{K^E_{U_1},0}\otimes\cdots\otimes\Phi_{K^E_{U_m},0})(\rho^{(m)})$, so the coherent information is $H(B^m)-H(E^m)$. By \cref{eq:degrade}, the state of $E^m$ under $(U_i)$ equals the state of $B_1^m$ under $(SU_i)$. Left multiplication by $S$ preserves the Haar measure, so $\int\dd U^m\,H(E^m)=\int\dd U^m\,H(B_1^m)$, and subtracting gives the equality in \cref{eq:averaged}. For the inequality, let $A$ purify $\rho^{(m)}$. Then $2\Ic=I(A;B^m)-I(A;E^m)$, and \cref{eq:degrade} holds with the reference attached, so $\int\dd U^m\,I(A;E^m)=\int\dd U^m\,I(A;B_1^m)\le\int\dd U^m\,I(A;B^m)$ by monotonicity of mutual information.
\end{proof}

\begin{lemma}[Tangent bound]
\label{lem:tangent}
Let $R\ge N/2$ and fix $P>0$. There is a number $y=y(P)\ge0$ such that, for every $m$ and every state $\rho^{(m)}$ of $mT$ modes,
\begin{equation}
\int\dd U^m\,\Ic\parens[\Big]{\rho^{(m)},\bigotimes_{i=1}^m\Phi_{U_i}}\le m\,\bar I_N(P,0)+y\,\squ[\big]{\meanN{\rho^{(m)}}-mP}.
\label{eq:tangent}
\end{equation}
\end{lemma}

The lemma adapts the relative-entropy argument of Theorem~6 of Ref.~\cite{WildeQi2018}, in which the logarithm of a Gibbs reference state turns an entropy difference into an affine function of the energy. Here the Haar average plays the role of the symmetry that reduces the affine function to the photon number alone.

\begin{proof}
Let $\sigma_U=\Phi_U(\isoinput{P})$, so that $\bar I_N(P,0)=\int\dd U\,\Ic(\isoinput{P},\Phi_U)$ by \cref{thm:singleuse} and \cref{eq:JN}. The proof has three steps.

\textit{Step 1: one relative-entropy inequality.} Let $\omega=(\bigotimes_i\Phi_{U_i})(\rho^{(m)})$ and $\sigma=\bigotimes_i\sigma_{U_i}$. The support of $\omega$ lies in the support of $\sigma$. Indeed, $\sigma_U$ is a gauge-invariant Gaussian state, faithful on the output directions in the range of $K_U$. Every output direction $w$ with $K_U^\dagger w=0$ is in the vacuum and in product with the rest, for every input, by \cref{eq:charfun}. The same holds for $\Tr_{B_2}\sigma_U$ and $\Tr_{B_2}\Phi_U(\rho)$ by \cref{lem:charfun}(c). Relative entropy does not increase under the partial trace over $B_2^m$~\cite{Lindblad1975}. Writing $D(\omega\Vert\sigma)=-H(\omega)-\Tr\omega\log\sigma$ on the support, this gives
\begin{equation}
H(B_2^m|B_1^m)_\omega\le-\Tr\omega\log\sigma+\Tr\omega_{B_1^m}\log\sigma_{B_1^m}.
\end{equation}
Both $\sigma$ and $\sigma_{B_1^m}$ are products over the $m$ uses, so their logarithms are sums of single-use terms, and only the marginals $\Phi_{U_i}(\rho_i)$ of $\omega$ enter, where $\rho_i$ is the $i$th marginal of $\rho^{(m)}$. Hence
\begin{align}
H(B_2^m|B_1^m)_\omega&\le\sum_{i=1}^mL_{U_i}(\rho_i),\label{eq:Lsum}\\
L_U(\rho)&=-\Tr\squ[\big]{\Phi_U(\rho)\log\sigma_U}\notag\\
&\quad+\Tr\squ[\big]{\parens{\Tr_{B_2}\Phi_U(\rho)}\log\Tr_{B_2}\sigma_U}.\notag
\end{align}
For $\rho^{(m)}=\isoinput{P}^{\otimes m}$ both relative entropies vanish, so equality holds and $L_U(\isoinput{P})=H(B_2|B_1)_{\sigma_U}$.

\textit{Step 2: the bound is affine in the second moments.} The state $\sigma_U$ is gauge invariant and Gaussian, so on its support $\log\sigma_U$ is a number-conserving quadratic form $-\sum_{jk}\beta_{jk}\hat a_j^\dagger\hat a_k$ in the output modes, plus a constant. The same holds for $\Tr_{B_2}\sigma_U$. Thus $L_U(\rho)$ is an affine function of the moments $\Tr[\Phi_U(\rho)\hat a_j^\dagger\hat a_k]$ and $\Tr[(\Tr_{B_2}\Phi_U(\rho))\hat a_j^\dagger\hat a_k]$. With a vacuum environment, these are linear in the input moments $C_\rho=\parens{\Tr[\rho\,\hat a_k^\dagger\hat a_j]}_{jk}$ by \cref{eq:charfun}. Therefore $L_U(\rho)=\Tr[Y_UC_\rho]+\gamma_U$ for a Hermitian $T\times T$ matrix $Y_U$ and a constant $\gamma_U$. All traces are finite because the input has finite photon number.

To justify Haar averaging, set $n=P/T>0$, $A_U=K_U^\dagger K_U$, and $A_{1,U}=K_{1,U}^\dagger K_{1,U}$, where $K_{1,U}$ contains the first $N-R$ rows of $K_U$. The thermal logarithms give
\begin{align*}
Y_U&=f_n(A_U)-f_n(A_{1,U}),\\
\gamma_U&=\Tr\log(I+nA_U)-\Tr\log(I+nA_{1,U}),
\end{align*}
where $f_n(0)=0$ and $f_n(x)=x\log(1+1/(nx))=(\ln2)^{-1}\int_0^{1/n}x/(x+s)\,\dd s$ for $x>0$. Inversion reverses the positive-semidefinite order, so $A\mapsto A(A+sI)^{-1}=I-s(A+sI)^{-1}$ is increasing for $s>0$. Since $0\le A_{1,U}\le A_U\le I$, integration and eigenvalue monotonicity give
\begin{equation*}
0\le Y_U\le\log(1+1/n)I,\qquad 0\le\gamma_U\le T\log(1+n).
\end{equation*}
The coefficients are continuous at rank drops and uniformly bounded, so Haar averaging is justified.

\textit{Step 3: Haar averaging leaves only the photon number.} For $V\in U(T)$, \cref{lem:charfun}(a) gives $\Phi_{K_U,0}\circ\Phi_{V,0}=\Phi_{K_UV,0}=\Phi_{K_{U(V\oplus I)},0}$, and $\isoinput{P}$ is fixed by $\Phi_{V,0}$. Hence $\sigma_{U(V\oplus I)}=\sigma_U$ and $L_{U(V\oplus I)}(\rho)=L_U(\Ucal_V\rho\,\Ucal_V^\dagger)$. Right multiplication by $V\oplus I$ preserves the Haar measure, so $\bar L(\rho)=\int\dd U\,L_U(\rho)$ is invariant under $\rho\mapsto\Ucal_V\rho\,\Ucal_V^\dagger$ for all $V\in U(T)$. This transformation sends $C_\rho$ to $VC_\rho V^\dagger$, so $\bar Y=\int\dd U\,Y_U$ commutes with all of $U(T)$ and equals $yI$ with $y\ge0$, since $Y_U\ge0$. This is the analogue of the symmetry argument in Telatar's theorem~\cite{Telatar1999}, which Ref.~\cite{DiazPerezAbreu2017} extends to channels with block distributional symmetries. Consequently $\bar L(\rho)=y\Tr C_\rho+\bar\gamma=y\meanN\rho+\bar\gamma$. Evaluating at $\isoinput{P}$ with \cref{lem:averaged} gives $\bar L(\isoinput{P})=\int\dd U\,H(B_2|B_1)_{\sigma_U}=\bar I_N(P,0)$, so $\bar\gamma=\bar I_N(P,0)-yP$. Averaging \cref{eq:Lsum} over $U^m$, using \cref{lem:averaged} on the left and $\sum_i\meanN{\rho_i}=\meanN{\rho^{(m)}}$ on the right, yields \cref{eq:tangent}.

\end{proof}

Two consequences are worth recording. The isotropic thermal state of photon number $P$ is optimal among all inputs of photon number at most $P$, since \cref{eq:tangent} with $m=1$ and $y\ge0$ gives $\int\dd U\,\Ic(\rho,\Phi_U)\le \bar I_N(P,0)$. Moreover, $\bar I_N(\cdot,0)$ is concave, because it lies below a line touching it at every $P$. It is also nondecreasing, because these lines have nonnegative slopes.

\begin{proof}[Proof of \cref{thm:capacity}(ii)]
\textit{Upper bound for $R\ge N/2$.} Take a fading code with error $\varepsilon$ and photon number at most $mP$. Let $\sigma=(\id_C\otimes\mc E)(\Fme_{CC'})$, let $\omega_{U^m}=(\id_C\otimes\bigotimes_i\Phi_{U_i})(\sigma)$, let $\hat\omega_{U^m}=(\id_C\otimes\mc D_{U^m})(\omega_{U^m})$, and let $\bar{\hat\omega}=\int\dd U^m\,\hat\omega_{U^m}$. The argument has five steps.

First, the fidelity with the pure state $\Fme_{C\hat B}$ is linear, so $F(\Fme_{C\hat B},\bar{\hat\omega})\ge1-\varepsilon$ and $\frac12\norm{\bar{\hat\omega}-\Fme_{C\hat B}}_1\le\sqrt\varepsilon$~\cite{FuchsVanDeGraaf1999}. The systems $C$ and $\hat B$ are $d$-dimensional, so the continuity bound for conditional entropy~\cite{AlickiFannes2004,Winter2016} gives
\begin{equation}
\log d\le I(C\rangle\hat B)_{\bar{\hat\omega}}+2\sqrt\varepsilon\log d+(1+\sqrt\varepsilon)\,h_2\parens[\Big]{\tfrac{\sqrt\varepsilon}{1+\sqrt\varepsilon}},
\end{equation}
where $I(C\rangle\hat B)=-H(C|\hat B)$ and $h_2$ is the binary entropy.

Second, conditional entropy is concave, so $I(C\rangle\hat B)_{\bar{\hat\omega}}\le\int\dd U^m\,I(C\rangle\hat B)_{\hat\omega_{U^m}}$. This step moves the average over $U^m$ inside without forming a flagged state.

Third, the decoder acts on $B^m$ alone, so data processing gives $I(C\rangle\hat B)_{\hat\omega_{U^m}}\le I(C\rangle B^m)_{\omega_{U^m}}$.

Fourth, decompose $\sigma=\sum_xq_x\phi^x$ into pure states with marginals $\rho^x$ on the $mT$ input modes. Concavity of conditional entropy and \cref{eq:Ic} give
\begin{equation}
\int\dd U^m\,I(C\rangle B^m)_{\omega_{U^m}}\le\sum_xq_x\int\dd U^m\,\Ic\parens[\Big]{\rho^x,\bigotimes_i\Phi_{U_i}},
\end{equation}
since $C$ purifies each $\rho^x$. The photon numbers satisfy $\sum_xq_x\meanN{\rho^x}=\meanN{\mc E(\pi_d)}\le mP$.

Fifth, for $P>0$ (the case $P=0$ is trivial), \cref{lem:tangent} bounds each term by $m\,\bar I_N(P,0)+y[\meanN{\rho^x}-mP]$. Averaging over $x$ with $y\ge0$ gives at most $m\,\bar I_N(P,0)$. The bound is affine in the photon number, so no concavity bookkeeping is needed.

Dividing by $m$ and letting $\varepsilon\to0$ gives $Q^F_P\le \bar I_N(P,0)$. Part (i) gives the reverse inequality.

\textit{The case $R\le N/2$.} Exchange the roles of $B$ and $E$ in \cref{lem:degrade}, with $S$ now exchanging output modes $1,\dots,R$ with $R+1,\dots,2R$ and $E=E_1E_2$ split after the first $R$ environment modes. Then $\Tr_{E_2}\circ\,\Phi_{K^E_U,0}=\Phi_{SU}$, and with $A$ purifying $\rho^{(m)}$,
\begin{equation}
2\int\dd U^m\,\Ic\parens[\Big]{\rho^{(m)},\bigotimes_i\Phi_{U_i}}=\int\dd U^m\squ[\big]{I(A;B^m)-I(A;E^m)}\le0,
\end{equation}
because $\int\dd U^m\,I(A;B^m)=\int\dd U^m\,I(A;E_1^m)\le\int\dd U^m\,I(A;E^m)$. The first four steps above then give $Q^F_P\le0$. A majority threshold of the same kind appears in quantum secret sharing through almost every passive interferometer~\cite{ArzaniEtAl2019}.
\end{proof}

\begin{remark}[Alternative route]
\Cref{lem:tangent} replaces a four-step argument in the style of Telatar's proof. One first twirls the input over the finite group of generalized permutation matrices with nonzero entries in $\{\pm1,\pm i\}$. The twirl fixes the photon number and produces the first and second moments of $\isoinput{P}$, and concavity of conditional entropy ensures that it does not lower the averaged coherent information. One then passes to $\isoinput{P}$ by Gaussian extremality of conditional entropy~\cite{WolfGiedkeCirac2006}, single-letterizes by strong subadditivity, and finally uses concavity and nonnegativity of $\bar I_N(\cdot,0)$ to replace the photon-number constraint by an equality. The tangent bound does all four steps at once.
\end{remark}

\section*{Acknowledgment}

Parts of this manuscript were prepared with the assistance of ChatGPT 5.6, developed
by OpenAI, and Claude Code, developed by Anthropic.
These agents were used to identify related literature, simplify the proof of the thermal fading capacity, and edit the writing of some sections, in accordance with the authors’ instructions.
All content, claims,
and conclusions have been reviewed and verified by the
authors to ensure accuracy and originality. Certain products, commercial and otherwise, are mentioned in this
publication. These mentions are for informational purposes only, and do not imply recommendation or endorsement by NIST.

\bibliographystyle{IEEEtran}
\bibliography{refs}

\end{document}